\documentclass[a4paper,11pt]{article}
\usepackage{jheppub} 
\usepackage{xcolor}
\usepackage{physics}
\usepackage{xspace}
\usepackage{mathrsfs}
\usepackage{soul}
\usepackage{mathtools}
\usepackage{tensor}
\usepackage{tabularx}
\usepackage{cancel}
\usepackage{booktabs}
\usepackage{marginnote}

\makeatletter
\g@addto@macro\bfseries{\boldmath}
\makeatother

\newcommand{\M}{\mathcal{M}}
\newcommand{\wedgedot}{\dot{\wedge}}
\newcommand{\ggamma}{\mathbf{\Gamma}}
\newcommand{\diag}{\text{diag}}
\newcommand{\gaugefield}{C}
\newcommand{\gaugefieldtilde}{\tilde{C}}
\newcommand{\gaugefieldhat}{\hat{C}}

\preprint{{Imperial-TP-2024-CH-5}\\ \rightline{{UUITP-24/24}}}
\title{Generalised symmetries in linear gravity}

\author[a]{Chris Hull,}
\author[a]{Maxwell L. Hutt,}
\author[a,b]{Ulf Lindstr\"{o}m}

\affiliation[a]{The Blackett Laboratory, Imperial College London, Prince Consort Road, London, SW7 2AZ, UK}
\affiliation[b]{Department of Physics and Astronomy, Uppsala University, Box 516, SE-75120 Uppsala, Sweden and Centre for Geometry and Physics, Uppsala University, Box 480, SE-75106 Uppsala, Sweden}

\emailAdd{c.hull@imperial.ac.uk, m.hutt22@imperial.ac.uk, ulf.lindstrom@physics.uu.se}

\abstract{
Linearised gravity has a global symmetry under which the graviton is shifted by a symmetric tensor satisfying a certain flatness condition. 
There is also a dual symmetry that can be associated with a global shift symmetry of the dual graviton theory.
The corresponding conserved charges are shown to satisfy a centrally-extended algebra. We discuss the gauging of these global symmetries, finding an obstruction to the simultaneous gauging of both symmetries which we interpret as a mixed 't Hooft anomaly for the ungauged theory. We discuss the implications of this, analogous to those resulting from a similar structure in Maxwell theory, and interpret the graviton and dual graviton as Nambu-Goldstone modes for these shift symmetries. 
}

\begin{document}
\maketitle
\flushbottom

\section{Introduction}
\label{sec:intro}

Generalised symmetries have been useful for understanding the infrared (IR) behaviour of quantum field theories (QFTs).
Different phases of gauge theories are distinguished according to which of their global symmetries are spontaneously broken; this is often referred to as the `Landau paradigm'.
The symmetry breaking pattern of
generalised global symmetries has provided an important understanding of the phase structure of certain theories (see, e.g. \cite{McGreevy:2022oyu, Bhardwaj:2023fca, Bhardwaj:2024wlr} and references therein).
For example, an unbroken 1-form symmetry signals a confining phase in certain theories \cite{Gaiotto2015GeneralizedSymmetries}.

The application of the paradigm to higher-form global symmetries is nicely illustrated by Maxwell theory in $d$ dimensions. This has both an `electric' $U(1)$ 1-form symmetry and a `magnetic' $U(1)$ $(d-3)$-form symmetry. The former acts on Wilson loops while the latter acts on 't~Hooft operators, measuring their electric and magnetic charges respectively. A Coulomb phase is characterised by the spontaneous breaking of both of these symmetries, in which both large Wilson and large 't Hooft loops pick up a non-zero expectation value \cite{Gaiotto2015GeneralizedSymmetries}. Higher-form generalisations of Goldstone's theorem \cite{Hofman:2018lfz, Lake:2018dqm} then imply the existence of a massless mode, which can be identified as the photon. Indeed, the photon can be understood as a Nambu-Goldstone boson for the  shift symmetry under which $A\to A+\lambda$ (where $\lambda$ is a flat connection).
This gives a satisfying explanation of the masslessness of the photon in terms of the breaking of a \emph{global} symmetry.
The magnetic current generates a symmetry of the dual formulation, written in terms of a $(d-3)$-form connection $\tilde{A}$, where it acts as a shift symmetry $\tilde{A} \to \tilde{A} + \tilde{\lambda}$ (with $\tilde{\lambda}$ a flat $(d-3)$-form connection). The dual photon can be understood similarly as the Nambu-Goldstone associated with the spontaneous breaking of this symmetry.

Another key feature of the higher-form symmetries of Maxwell theory is their mixed 't Hooft anomaly.
Such anomalies are of great importance as they are preserved along renormalisation group (RG) flow, and therefore must be matched in the IR \cite{tHooft:1979rat}. 
This places constraints on the low-energy dynamics of theories with such an anomaly, the classic example of which is the result of Lieb, Schultz and Mattis \cite{Lieb:1961fr} showing that certain spin chains cannot have a trivial gapped ground state.
More recently, similar arguments involving anomaly matching for mixed 't Hooft anomalies of discrete and generalised symmetries have shed light on the IR structure of many gauge theories, including pure Yang-Mills (in particular at $\theta=\pi$) and QCD (see, e.g.\  \cite{Gaiotto:2017yup, Cordova:2019bsd, Cordova:2019jnf, Cordova:2019uob, Brennan:2024tlw, Brennan:2023ynm, Apte:2022xtu, Cordova:2019jqi}). 

A novel perspective was provided in \cite{Delacretaz:2019brr}, where it was shown that a particular mixed 't Hooft anomaly between a pair of global higher-form $U(1)$ symmetries can be responsible for the existence of a gapless mode, without explicit reference to spontaneous symmetry breaking. Indeed, this reasoning is readily applied to the mixed anomaly of Maxwell theory, giving another explanation of the masslessness of the photon.\footnote{As in \cite{Delacretaz:2019brr, Hinterbichler2023GravitySymmetries}, we will refer to excitations whose masslessness is protected by a 't Hooft anomaly as Nambu-Goldstone modes, even when there is no explicit reference to spontaneous symmetry breaking.}

A useful approach to the mixed anomaly is to consider the \emph{gauging} of the electric and magnetic generalised symmetries. This involves attempting to promote these global symmetries to local ones by coupling to suitable gauge fields.
The mixed 't Hooft anomaly in the global symmetries manifests itself as an obstruction to simultaneously gauging the electric and magnetic symmetries, and finding this obstruction is a convenient way of calculating the anomaly \cite{Harvey:2005it}.

Recently, there has been substantial interest in generalised symmetries in gravity \cite{Hinterbichler2023GravitySymmetries, Benedetti2022GeneralizedGraviton, Benedetti2023GeneralizedGravitons, BenedettiNoether, Cheung:2024ypq, Gomez-Fayren2023CovariantRelativity, Hull:2024mfb, Hull:2024xgo}, and in particular in the linearised theory.
It has been argued that the linearised theory can emerge as the IR fixed point of gravity coupled to massive fields \cite{Farnsworth:2021zgj}, so one would hope that understanding its symmetries could give some insight into the non-linear theory.

In \cite{Hinterbichler2023GravitySymmetries}, arguments similar to those outlined above were made for gravitons moving on a flat background spacetime.
It was found that a mixed 't~Hooft anomaly between a pair of electric and magnetic `bi-form symmetries' is responsible for the masslessness of the graviton. 
In that work, the electric symmetry used was a graviton shift of the form $h_{\mu\nu} \to h_{\mu\nu} + \partial^\rho \Lambda_{\rho\mu|\nu}$.
This has the virtue of having an associated Noether current that is gauge-invariant but gives a rather different set-up than the one for the photon, and in particular it is not so natural to regard the graviton as a Nambu-Goldstone boson for this symmetry. 

Here we consider instead the global symmetries consisting of general shifts of the graviton and of the dual graviton.
These are global symmetries for restricted parameters\footnote{These restrictions on the parameters are analogous to the flatness condition $\dd\lambda=0$ on the shift of the photon $A\to A+\lambda$.}
and we consider the gauging in which these are promoted to local symmetries.
We find a significantly simpler structure for the gauging than was found for the shift considered in \cite{Hinterbichler2023GravitySymmetries}.
The non-linearities of Einstein gravity explicitly break these symmetries and so if the linearised theory arose as an IR limit \cite{Farnsworth:2021zgj} they would be emergent symmetries.
A key role in our analysis is played by gravitational duality \cite{Hull2000}: the free graviton theory has a dual formulation in terms of a dual graviton field, which is a bi-form,\footnote{A $[p,q]$ bi-form is in the tensor product of the space of $p$-forms with the space of $q$-forms; see section \ref{sec:biforms}. The dual graviton in $d$ dimensions is a $[d-3,1]$ bi-form gauge field.} and shifts of this field are a bi-form symmetry of the dual formulation.

As for the photon, we find  a mixed 't Hooft anomaly for these two shift symmetries which can be understood in terms of an obstruction to simultaneously gauging them.
This anomaly is closely related to the anomaly discussed in \cite{Hinterbichler2023GravitySymmetries}. In particular, the anomalous Ward identities which result from this mixed anomaly can be shown to imply anomalous Ward identities similar in form to those studied in \cite{Hinterbichler2023GravitySymmetries}. These in turn imply the masslessness of the graviton.

The significant simplifications that result from considering these shift symmetries rather than the ones discussed in \cite{Hinterbichler2023GravitySymmetries} make the generalised symmetry structure and gauging considerably more transparent. In particular, it is natural to regard the graviton as the Nambu-Goldstone boson for the electric shift symmetry and the dual graviton as that for the magnetic shift symmetry.
However, in our case, the associated Noether currents are \emph{not} invariant under the graviton gauge symmetry $h_{\mu\nu} \to h_{\mu\nu} + 2\partial_{(\mu} \xi_{\nu)}$ (or the dual graviton gauge symmetry in the dual formulation).
The lack of gauge invariance means that these currents do not correspond to  well-defined local operators in the quantum theory, but gauge-invariant currents can be constructed from derivatives of these currents and these are local observables.
The correlation functions of these observables are readily calculated from the correlation functions of our non-gauge-invariant currents. 

Since the Noether currents associated with the global symmetries we consider are not fully anti-symmetric tensors, they cannot directly be integrated over submanifolds. However, gauge-invariant higher-dimensional topological operators
can be constructed from our non-gauge-invariant currents. This is done
by first contracting the current with a suitable background tensor
to make a current that is a differential form which can then be integrated over a suitable surface to give a topological operator.\footnote{This situation is not unfamiliar. For example, the $\mathbb{Z}_k$ 1-form symmetry in $U(1)_k$ Chern-Simons theory is generated by topological Wilson lines \cite{Gaiotto2015GeneralizedSymmetries}, which are also a gauge-invariant integral of a non-gauge-invariant local density. The same is true of the topological defects in the well-known non-invertible construction of \cite{Choi:2022jqy}.}
These operators then correspond to a set of (higher-form) symmetries.

The remainder of this paper is set out as follows. In section~\ref{sec:maxwell}, we review the higher-form symmetries of Maxwell theory and their mixed 't Hooft anomaly. 
In section~\ref{sec:graviton_symmetries} we consider the shift symmetry of the graviton and its gauging, as well as a dual symmetry.
The dual graviton theory and its symmetries are studied in section~\ref{sec:shifts_for_dual}, with the result that the dual symmetry of the graviton can be interpreted as a shift symmetry of the dual graviton (see section~\ref{sec:relating_dual_theories}).
The mixed 't Hooft anomaly between these two symmetries is described in section~\ref{sec:anomalies}. In section~\ref{sec:massless_graviton}, we discuss the implications of  this anomaly and its relation to the results of \cite{Hinterbichler2023GravitySymmetries}.
Finally, in section \ref{sec:discussion}, we make some remarks about observables in the graviton theory and discuss further implications of our results.

\section{Higher-form symmetries of Maxwell theory}
\label{sec:maxwell}

Our treatment of linear gravity will follow and generalise that for spin-1, so we begin with a brief review of Maxwell theory on a $d$-dimensional manifold $\mathcal{M}$ with Lorentzian signature. The curvature 2-form $F$ can be written locally in terms of a $U(1)$ connection 1-form $A$, with $F=\dd{A}$. The action is
\begin{equation}\label{eq:S_Maxwell}
    S_{\text{M}} = -\frac{1}{2e^2} \int_{\mathcal{M}} F \wedge \star F
\end{equation}
where $e$ is a coupling constant.
The Bianchi identity $\dd{F} = 0$ and the field equation $\dd\star F=0$ imply the existence of two conserved 2-form currents
\begin{equation}
    J_{\text{e}} = \frac{1}{e^2} F\qc J_{\text{m}} = \frac{1}{2\pi} \star F
\end{equation}
These currents can  be integrated to give
 topological operators 
\begin{equation}\label{eq:Q_m}
    Q_{\text{e}} = \int_{\Sigma_{d-2}} \star J_e\qc Q_{\text{m}} = \int_{\Sigma_2} \star J_m
\end{equation}
for any codimension-2 cycle $\Sigma_{d-2}$ and dimension-2 cycle $\Sigma_2$.
The charge $Q_{\text{e}}$ generates a $U(1)$ symmetry acting on Wilson lines and the charge  $Q_{\text{m}}$ generates a $U(1)$ symmetry acting on 't Hooft lines in $d=4$, or 't Hooft $(d-3)$-branes in $d>4$ \cite{Gaiotto2015GeneralizedSymmetries}.

The field strength $F$ is invariant under shifts of the gauge field $A\to A+\lambda$ by a 1-form $\lambda$ which is closed, $\dd\lambda=0$. 
This is then a symmetry of the action associated with the conserved current $J_{\text{e}}$.
The global symmetries of this type are those which are not gauge symmetries, which means that $\lambda $ is closed but not exact, so that the symmetry group is given by the first cohomology group.
In fact, the symmetry is slightly larger: the gauge field can be shifted by a flat connection modulo gauge transformations \cite{Gaiotto2015GeneralizedSymmetries}.

The theory can also be formulated in terms of a dual $(d-3)$-form potential $\tilde{A}$ with $F=\star \dd \tilde{A}$. There is then a $(d-3)$-form shift symmetry $\tilde{A}\to \tilde{A} + \tilde{\lambda}$ with $\dd\tilde \lambda=0$ associated with the conserved current $J_{\text{m}}$. 
Again, the non-trivial transformations correspond to closed forms $\tilde{\lambda}$ modulo exact ones.
As before, there is a larger symmetry in which $\tilde{A}$ is shifted by a flat gerbe connection modulo gauge transformations.

Note that, in the standard formulation of the theory in terms of the 1-form gauge field $A$, the `magnetic' $(d-3)$-form symmetry does not correspond to any transformation of the gauge field $A$ and in the dual description in terms of the $(d-3)$-form gauge field $\tilde{A}$, the `electric' 1-form symmetry does not correspond to a transformation of the gauge field $\tilde{A}$.

These two higher-form symmetries are viewed as global symmetries, as the parameters are restricted to be closed, and have a mixed 't Hooft anomaly \cite{Gaiotto2015GeneralizedSymmetries}. This can be seen by coupling  to background gauge fields. The electric 1-form symmetry can be gauged by introducing a 2-form background $U(1)$ gauge field $B$ which transforms as $B \to B + \dd\lambda$. This implies that the combination $F-B$ is invariant, and the gauging is achieved by replacing $F$ in \eqref{eq:S_Maxwell} with $F-B$,
\begin{equation}\label{eq:Maxwell_electric_gauged}
    S_{\text{M}}^{\text{e}} = -\frac{1}{2e^2} \int_{\mathcal{M}} (F-B)\wedge\star(F-B)
\end{equation}
 Since the magnetic symmetry is not related to any transformation of the gauge field $A$, it can be gauged by introducing the standard coupling of the current $J_{\text{m}}$ to a background $(d-2)$-form $U(1)$ gauge field $\tilde{B}$,
\begin{equation}
    S_{\text{M}}^{\text{m}} = -\frac{1}{2e^2} \int_{\mathcal{M}} F\wedge\star F + \frac{1}{2\pi} \int_{\mathcal{M}} F \wedge \tilde{B}
\end{equation}
where $\tilde{B} \to \tilde{B} + \dd{\tilde{\lambda}}$ under the magnetic symmetry.

Introducing both the electric and magnetic background fields results in
\begin{equation}
    S_{\text{M}}^{\text{e}+\text{m}} = -\frac{1}{2e^2} \int_{\mathcal{M}} (F-B)\wedge\star(F-B) + \frac{1}{2\pi} \int_{\mathcal{M}} F \wedge \tilde{B}
\end{equation}
This is invariant under the magnetic transformation $\tilde{B} \to \tilde{B} + \dd{\tilde{\lambda}}$, but is no longer invariant under the electric transformations $A\to A+\lambda$, $B\to B+\dd{\lambda}$, signalling an anomaly. In particular, we find
\begin{equation}\label{eq:Maxwell_anomaly}
    \delta S_{\text{M}}^{\text{e}+\text{m}} = \frac{1}{2\pi} \int_{\mathcal{M}} \dd{\lambda} \wedge \tilde{B}
\end{equation}
A counterterm $-\frac{1}{2\pi} B\wedge\tilde{B}$ can be added to the Lagrangian to cancel this variation, but the resulting action is then not invariant under the magnetic transformation. 
There is no counterterm that gives an action that is invariant under both symmetries. 

The anomaly can be cancelled by the inflow mechanism if it is coupled to a $(d+1)$-dimensional theory. This involves extending $B$ and $\tilde{B}$ into a $(d+1)$-dimensional bulk $\mathcal{N}$, with $\partial\mathcal{N} = \mathcal{M}$, and coupling to a topological field theory with a BF action
\begin{equation}\label{eq:S_inflow_Maxwell}
    S_{\text{M}}^{\text{inflow}} = -\frac{1}{2\pi} \int_{\mathcal{N}} B \wedge \dd\tilde{B}
\end{equation}
This $(d+1)$-dimensional theory is non-trivial, indicating that there is no choice of $d$-dimensional counterterm which will remove the anomaly. 
Alternatively, we can understand the anomaly via the descent procedure. The $(d+2)$-form anomaly polynomial which produces the $(d+1)$-dimensional  action \eqref{eq:S_inflow_Maxwell} by descent is 
\begin{equation}\label{eq:anomaly_poly_maxwell}
    \mathcal{I}_{d+2}^{\text{M}} = -\frac{1}{2\pi} \dd{B} \wedge \dd{\tilde{B}}
\end{equation}

A mixed 't Hooft anomaly between two symmetries implies that gauging one explicitly breaks the other. In terms of the topological operators $Q_\text{e}$ and $Q_\text{m}$, this means that gauging the electric symmetry, for example, will render the magnetic operator $Q_\text{m}$ non-topological. 
While the Bianchi identity $\dd{F}=0$ remains valid in the gauged theory, $F$ is no longer a gauge-invariant operator. The gauge-invariant quantity of interest is $F-B$, which is not closed for generic background field configurations since $\dd{(F-B)} = -\dd{B} \neq0$. Therefore, $Q_m$ as defined in \eqref{eq:Q_m} is not a gauge-invariant operator in the gauged theory, and modifying its definition to $Q_m = \frac{1}{2\pi} \int_{\Sigma_2} (F-B)$ results in an operator which is gauge-invariant but not topological. The magnetic symmetry is, therefore, broken when the electric symmetry is coupled to a background gauge field. 

For further discussion of the generalised symmetries and anomaly in Maxwell theory, see e.g.\ the reviews in \cite{Hofman:2017vwr, Brennan:2023mmt, Schafer-Nameki:2023jdn, Bhardwaj:2023kri} and references therein.
\section{Bi-form calculus}
\label{sec:biforms}

In this section we review the bi-form calculus of \cite{Medeiros2003ExoticDuality} that will be useful in what follows and establish our notation. 
A $[p,q]$ bi-form is a tensor which is an element of $\Omega^p \otimes \Omega^q$, where $\Omega^p$ is the space of $p$-forms on a $d$-dimensional manifold $\M$. We also refer to such objects as $[p,q]$-tensors. A $[p,q]$ bi-form has index symmetries
\begin{equation}
    A_{\mu_1\dots\mu_p|\nu_1\dots\nu_q} = A_{[\mu_1\dots\mu_p]|\nu_1\dots\nu_q} = A_{\mu_1\dots\mu_p|[\nu_1\dots\nu_q]}
\end{equation}
Those bi-forms with $p>q$ which further satisfy
\begin{equation}
    A_{[\mu_1\dots\mu_p|\nu_1]\nu_2\dots\nu_q} = 0
\end{equation}
are irreducible under $GL(d,\mathbb{R})$. They transform in the $GL(d,\mathbb{R})$ representation labelled by a Young tableau with two columns of lengths $p$ and $q$.
For $p=q$, the extra  condition
\begin{equation}
   A_{\mu_1\dots\mu_p|\nu_1\dots\nu_q}= A_{\nu_1\dots\nu_q|\mu_1\dots\mu_p}
\end{equation}is needed.
Gauge fields in general $GL(d,\mathbb{R})$ representations  were introduced by Curtright \cite{Curtright:1980yk} and have
been studied in e.g.\ \cite{Medeiros2003ExoticDuality, Bekaert2004, Dubois-Violette:1999iqe,Dubois-Violette:2000fok, Dubois-Violette:2001wjr,Howe2018SCKYT,  deMedeiros:2003osq, Hull2001DualityFields, Henneaux:2004jw, Hinterbichler2023GravitySymmetries, Hinterbichler:2024cxn} and references therein. We will discuss only bi-form gauge fields here.

We will now briefly review some of the operations on bi-forms
introduced in \cite{Medeiros2003ExoticDuality}. There are left and right exterior derivatives\footnote{These were referred to as $\dd$ and $\tilde{\dd}$ in \cite{Medeiros2003ExoticDuality}.}
\begin{align}
\begin{split}
    \dd_L : \Omega^p \otimes \Omega^q &\to \Omega^{p+1} \otimes \Omega^q \\
    \dd_R : \Omega^p \otimes \Omega^q &\to \Omega^p \otimes \Omega^{q+1}
\end{split}
\end{align}
defined by
\begin{equation}
\begin{split}
    (\dd_L A)_{\mu_1\dots\mu_{p+1}|\nu_1\dots\nu_q} &= \partial_{[\mu_1} A_{\mu_2\dots\mu_{p+1}]|\nu_1\dots\nu_q} \\
    (\dd_R A)_{\mu_1\dots\mu_p|\nu_1\dots\nu_{q+1}} &= A_{\mu_2\dots\mu_{p+1}|[\nu_1\dots\nu_q,\nu_{q+1}]}
\end{split}
\end{equation}
where a comma denotes partial derivative.
These satisfy $\dd_L^2 = \dd_R^2 = 0$ and commute, $[\dd_L,\dd_R]=0$, so $(\dd_L+\dd_R)^3=0$.
We note that $\dd_R$ does not map irreducible $[p,q]$ bi-forms to irreducible $[p,q+1]$ bi-forms. However, $\dd_L$ does map irreducible $[p,q]$ bi-forms to irreducible $[p+1,q]$ bi-forms.

Given a bi-form in a reducible representation, its projection onto the irreducible $[p,q]$ representation can be achieved via the use of a Young projector $\mathcal{Y}_{[p,q]}$. For example, for an $n$-form $\phi$ (i.e. an $[n,0]$ bi-form), the component of $\dd_R\phi$ in the irreducible $[n,1]$ representation is given by
\begin{equation}
    \left(\mathcal{Y}_{[n,1]} (\dd_R \phi)\right)_{\mu_1\dots\mu_n|\nu} = \partial_\nu \phi_{\mu_1\dots\mu_n} - \partial_{[\nu} \phi_{\mu_1\dots\mu_n]}
\end{equation}

We also introduce left and right duality operators $\star_L$ and $\star_R$ by
\begin{align}
    (\star_L A)_{\mu_1\dots\mu_{d-p}|\nu_1\dots\nu_q} &= \frac{1}{p!} \epsilon_{\mu_1\dots\mu_{d-p}\alpha_1\dots\alpha_p} A\indices{^{\alpha_1\dots\alpha_p}_{|\nu_1\dots\nu_q}} \\
    (\star_R A)_{\mu_1\dots\mu_p|\nu_1\dots\nu_{d-q}} &= \frac{1}{q!} \epsilon_{\nu_1\dots\nu_{d-q}\alpha_1\dots\alpha_q} A\indices{_{\mu_1\dots\mu_p|}^{\alpha_1\dots\alpha_q}}
\end{align}
where indices are raised with the Minkowski metric $\eta_{\mu\nu} = \diag(-1,1,\dots,1)$. These satisfy
\begin{equation}
    \star_L^2 = (-1)^{p(d-p)+1} \qc \star_R^2 = (-1)^{q(d-q)+1} 
\end{equation}
when acting on a $[p,q]$ form.
We further define  derivatives $\dd_L^\dag$ and $\dd_R^\dag$ which act as
\begin{align}
    (\dd_L^\dag A)_{\mu_2\dots\mu_p|\nu_1\dots\nu_q} &= \partial^{\mu_1} A_{\mu_1\dots\mu_p|\nu_1\dots\nu_q} \\
    (\dd_R^\dag A)_{\mu_1\dots\mu_p|\nu_2\dots\nu_q} &= \partial^{\nu_1} A_{\mu_1\dots\mu_p|\nu_1\dots\nu_q}
\end{align}
These have the usual structure $\dd^\dag_L \propto \star_L \dd_L \star_L$ and similarly for $\dd^\dag_R$, which implies that $(\dd^\dag_L)^2=0$ and $(\dd^\dag_R)^2=0$.
We will sometimes refer to tensors satisfying $\dd_L^\dag A=0$ as being left-conserved, and those satisfying $\dd_R^\dag A=0$ as being right-conserved.
Finally, we introduce an operation 
\begin{equation}
    \wedgedot: (\Omega^{p_1} \otimes \Omega^q )\times (\Omega^{p_2} \otimes\Omega^q ) \to \Omega^{p_1+p_2}
\end{equation}
defined by
\begin{equation}\label{eq:wedgedot}
    (A \,\wedgedot\, B)_{\mu_1\dots\mu_p} = A\indices{_{[\mu_1\dots\mu_{p_1}|}^{\nu_1\dots\nu_q}} B_{\mu_{p_1+1} \dots \mu_p]|\nu_1\dots\nu_q}
\end{equation}
where $A$ and $B$ are bi-forms of degrees $[p_1,q]$ and $[p_2,q]$ respectively, and $p = p_1+p_2$. In other words, $\wedgedot$ acts as the wedge product on the first set of indices of the tensors, and as an inner product on the second set. 
\section{Symmetries of the graviton theory}
\label{sec:graviton_symmetries}

\subsection{Linearised gravity}

We study the graviton theory on Minkowski space $\mathcal{M} = \mathbb{R}^{1,d-1}$ (possibly with some points/ regions removed) and assume fall-off conditions sufficient for total derivative terms to be ignored. The Fierz-Pauli action can be written\footnote{The original Fierz-Pauli theory \cite{Fierz:1939ix} also contained a mass term for $h$ which we are setting to zero here.}
\begin{equation}\label{eq:S_FP}
    S_{\text{FP}} = \frac{1}{2} \int_{\mathcal{M}} \dd[d]{x} h_{\mu\nu} G(h)^{\mu\nu} 
\end{equation}
where
\begin{equation}\label{eq:G(h)}
    G(h)\indices{^\mu_\nu} = -3 \delta^{\mu\rho\alpha}_{\nu\sigma\beta} \partial_\rho \partial^\sigma h\indices{_\alpha^\beta} 
\end{equation}
is the linearised Einstein tensor, with $\delta^{\mu\rho\alpha}_{\nu\sigma\beta} = \delta^{[\mu}_{\nu} \delta^\rho_\sigma \delta^{\alpha]}_\beta$. 
The gauge symmetries of the graviton are linearised diffeomorphisms, acting as
\begin{equation}\label{eq:linear_diffeo}
    h_{\mu\nu} \to h_{\mu\nu} + 2\partial_{(\mu} \xi_{\nu)}
\end{equation}
under which \eqref{eq:S_FP} is invariant. The invariant curvature is the (linearised) Riemann tensor
\begin{equation}\label{eq:riemann}
    R(h)_{\mu\nu\rho\sigma} = \partial_\rho \Gamma(h)_{\mu\nu|\sigma} - \partial_\sigma \Gamma(h)_{\mu\nu|\rho}
\end{equation}
where   $\Gamma(h)$ is given by
\begin{equation}\label{eq:Gamma}
    \Gamma(h)_{\mu\nu|\rho} = \partial_{[\mu} h_{\nu]\rho}
\end{equation}
Alternatively, we can view $h$ as an irreducible $[1,1]$ bi-form and the above relations can be written as $\Gamma(h) = \dd_L h$ and $R(h) = -2\dd_L \dd_R h$.\footnote{Note that $R(h)$ is irreducible, i.e. $R(h)_{[\mu\nu\rho]\sigma}=0$, without the use of a Young projector.}
Varying the graviton in the   action leads to the equation of motion
\begin{equation}
    G(h)_{\mu\nu} = 0
\end{equation}

\subsection{Shift symmetries of the graviton}
\label{sec:graviton_shifts}

We now consider a shift of the graviton
\begin{equation}\label{eq:graviton_shift}
    \delta h_{\mu\nu} = \alpha_{\mu\nu}
\end{equation}
where $\alpha$ is a symmetric rank-2 tensor (equivalently, a [1,1] bi-form). Such transformations were considered in \cite{Hinterbichler2023GravitySymmetries}, where they were referred to as [1,1] bi-form symmetries. 
The resulting variation of the action can be written as
\begin{equation}\label{eq:FP_variation}
    \delta S_{\text{FP}} = \int_{\mathcal{M}} \dd[d]{x} h_{\mu\nu} G(\alpha)^{\mu\nu}
\end{equation}
which vanishes provided that $\alpha$ satisfies 
\begin{equation}\label{eq:G(alpha)=0}
    G(\alpha)_{\mu\nu} = 0
\end{equation}
For such restricted parameters $\alpha$ the transformation \eqref{eq:graviton_shift} can then be regarded as  a global symmetry of the theory. This is analogous to the constraint that $\dd\lambda=0$ in order for $A\to A+\lambda$ to be a global symmetry of Maxwell theory. 
Note that this condition is weaker than the condition for the curvature to be invariant $R(\alpha) =  0$ where
$R(\alpha) = -2 \dd_L \dd_R \alpha $.\footnote{If the  graviton theory arises as an effective theory, we should allow also for higher-derivative terms which are constructed from the curvature $R(h)$. In such cases, the shift symmetry  persists provided that $\alpha$ satisfies the stronger constraint $R(\alpha)  = 0$ \cite{Hinterbichler2023GravitySymmetries}.}

For an unconstrained parameter $\alpha$, the variation of the action \eqref{eq:FP_variation} can be written
\begin{equation}\label{eq:dS_FP}
    \delta S_{\text{FP}} = \int_{\mathcal{M}} \dd[d]{x} \partial_\mu \partial_\nu \alpha_{\alpha\beta} K(h)^{\nu\beta|\mu\alpha}
\end{equation}
where $K(h)$ is an irreducible $[2,2]$ bi-form defined by
\begin{equation}
    K(h)\indices{_{\mu\nu|}^{\rho\sigma}} = -3 \delta_{\mu\nu\alpha}^{\rho\sigma\beta} h\indices{^\alpha_\beta}
\end{equation}
It follows from \eqref{eq:dS_FP} 
that $\dd_L^\dag \dd_R^\dag K(h)=0$ on-shell. Indeed, we find
\begin{equation}\label{eq:dLdRK=0}
    \dd_L^\dag \dd_R^\dag K(h) = G(h) \doteq 0
\end{equation}
where $\doteq$ is an on-shell equality. This can be viewed as a 2-derivative extension of the standard Noether arguments.

From $K$ we can construct a conserved [2,1] bi-form current $J(h)_{\mu\nu|\rho}$ given by\footnote{There is a similar conserved current defined  by $\dd_L^\dag K(h)$. Since $K$ is irreducible, this is equal to $\dd_R^\dag K$ with the two antisymmetric sets of  indices interchanged.}
\begin{equation}\label{eq:J}
    J(h) = \dd_R^\dag K(h)
\end{equation}
whose components are given by
\begin{equation}\label{JGam}
    J(h)\indices{_{\mu\nu|}^\rho} = 3 \delta_{\mu\nu\alpha}^{\rho\sigma\beta} \Gamma(h)\indices{_{\sigma\beta|}^\alpha}
\end{equation}
It  follows from \eqref{eq:dLdRK=0} and \eqref{eq:J} that $J(h)$ is left-conserved (i.e. conserved on the first set of indices) on-shell,
\begin{equation}\label{eq:J(h)_conservations}
    \dd_L^\dag J(h) \doteq 0
\end{equation}
and right-conserved off-shell,
\begin{equation}\label{eq:J_right_conserved}
    \dd_R^\dag J(h) = 0
\end{equation}
This conserved current can be viewed as a Noether-like current
associated with the continuous global symmetry \eqref{eq:graviton_shift}.\footnote{The conservation of $J(h)$ is preserved by the addition of a term of the form $\dd^\dag_L \dd^\dag_R H$, where $H$ is an arbitrary [3,2] bi-form.
In \cite{Hull:2024xgo, Hull:2024mfb}, the freedom to add a co-exact piece was used to relate certain non-gauge-invariant 2-form currents to gauge-invariant ones.
For the current $J(h)\indices{_{\mu\nu|}^\rho}$, there is no choice of $H$ which will make the current gauge-invariant, so we set $H=0$.} 
This current is not gauge-invariant and transforms by
\begin{equation}
    \delta J(h)\indices{_{\mu\nu|}^\rho} = 3 \delta_{\mu\nu\alpha}^{\rho\beta\gamma} \partial^\alpha \partial_\beta \xi_\gamma
\end{equation}
under a linearised diffeomorphism \eqref{eq:linear_diffeo}.

The shift symmetry \eqref{eq:graviton_shift} is the gravitational analogue of the electric 1-form symmetry in Maxwell theory which shifts the photon. We will see in section~\ref{sec:dual_symmetries} that there is a dual symmetry which can be related to a shift of the dual graviton, analogous to the magnetic $(d-3)$-form symmetry of Maxwell theory.

Shifts for which $\alpha_{\mu\nu} = 2 \partial_{(\mu}\xi_{\nu)}$ are gauge transformations \eqref{eq:linear_diffeo}, so the non-trivial transformations are given by equivalence classes of $\alpha$ satisfying $G(\alpha)=0$ modulo those  of the form $\alpha_{\mu\nu} = 2 \partial_{(\mu}\xi_{\nu)}$ for some $\xi_\nu$.

In \cite{Hinterbichler2023GravitySymmetries}, the particular shift 
\begin{equation}\label{eq:alpha=divLambda}
    \alpha_{\mu\nu} = \partial^\rho \Lambda_{\rho\mu|\nu}
\end{equation}
was considered.
This has the property that, when $\Lambda$ is viewed as the transformation parameter, the associated $[2,2]$ Noether current can be taken as the linearised Riemann tensor.
While this has the benefit of producing a gauge-invariant Noether current, it is not the most general shift symmetry of the graviton. Namely, an $\alpha$ of the form \eqref{eq:alpha=divLambda} necessarily satisfies $\partial^\mu \alpha_{\mu\nu}=0$, whereas the most general global shift symmetry of the graviton theory is one where $\alpha$ only satisfies the weaker constraint $G(\alpha)_{\mu\nu}=0$.

Since this particular shift with parameter of the form \eqref{eq:alpha=divLambda} is a special case of the more general one \eqref{eq:graviton_shift} that we consider here, the on-shell conservation of the $[2,2]$ Noether current $R$ should follow from that of $J(h)$. 
Indeed, we note that the connection $\Gamma$ can be written in terms of $J$ as
\begin{equation}
    \Gamma(h)\indices{_{\mu\nu|}^\rho} = J(h)\indices{_{\mu\nu|}^\rho} - \frac{2}{d-2} \delta^{\rho}_{[\mu} J(h)\indices{_{\nu]\alpha|}^\alpha}
\end{equation}
and acting on this equation with $\dd_R$ gives
\begin{equation}\label{eq:X_def}
    \frac{1}{2} R(h)\indices{_{\mu\nu}^{\rho\sigma}} = \partial^{[\rho} J(h)\indices{_{\mu\nu|}^{\sigma]}} - \frac{2}{d-2} \partial^{[\rho}\delta^{\sigma]}_{[\mu} J(h)\indices{_{\nu]\alpha|}^\alpha}
\end{equation}
The on-shell left-conservation of $R$ then follows from that of $J$.

A conserved $p$-form current can be integrated over a p-cycle to give a topological operator, but a conserved bi-form current cannot be directly integrated. However, a bi-form current can be contracted with a suitable tensor to give a conserved $p$-form current which can then be integrated to give a topological operator. We now give examples of such topological operators generating 0-form and 1-form symmetries for linearised gravity and will discuss further examples in a forthcoming work.

Note that while $J(h)$ is not gauge-invariant -- and so is not a well-defined local operator in the gauge theory -- the topological operators which can be constructed in this way can be gauge-invariant and well-defined.

For a given $\alpha$, there is a codimension-1 topological operator which generates the 0-form symmetry under which the graviton field transforms as \eqref{eq:graviton_shift}.
This can be constructed in the same way as the charge generating the photon shift \cite{Hullprep}.
Let us choose a particular $\alpha$ satisfying \eqref{eq:G(alpha)=0} and consider the shift
\begin{equation}\label{eq:0-form-shift}
    h_{\mu\nu} \to h_{\mu\nu} + \epsilon \alpha_{\mu\nu}
\end{equation}
where $\epsilon$ is a constant 0-form parameter. 
Using standard arguments, we find the Noether current by considering a transformation with non-constant parameter $\epsilon$, under which the variation of the action is
\begin{equation}
    \delta S_{\text{FP}} = \int_{\mathcal{M}} \dd[d]{x} \epsilon \partial_\mu j(\alpha)^\mu
\end{equation}
where
\begin{equation}\label{eq:j(alpha)}
    j(\alpha)_\alpha = -3 \delta^{\nu\gamma\delta}_{\mu\alpha\beta} \left( \alpha\indices{^\mu_\nu} \partial_\gamma h\indices{^\beta_\delta} - \partial_\gamma \alpha\indices{^\mu_\nu} h\indices{^\beta_\delta} \right)
\end{equation}
Then $j(\alpha)$ is the 1-form Noether current associated with the shift by this particular $\alpha$. In the simpler case when $\alpha$ satisfies the stronger constraint $\dd_L\alpha=0$ (which implies that $G(\alpha)=0$), this reduces to $j(\alpha)_\mu = J_{\mu\nu|\rho} \alpha^{\nu\rho}$. Indeed, it is straightforward to verify that $\partial^\mu j(\alpha)_\mu \doteq 0$ on-shell.

The codimension-1 topological operator which generates this shift is then
\begin{equation}\label{eq:Q(alpha)}
    Q(\alpha) = \int_{\Sigma_{d-1}} \star j(\alpha)
\end{equation}
for some codimension-1 surface $\Sigma_{d-1}$ which we either take to be closed, or assume that the fields satisfy suitable boundary conditions so that we can neglect surface terms.
If $\alpha$ is taken to be of the form 
\begin{equation}\label{eq:alpha=dxi}
    \alpha_{\mu\nu} = 2\partial_{(\mu} \xi_{\nu)}
\end{equation}
then the current can be written
\begin{equation}
    j(\alpha)_\alpha = 3\partial^\mu \left( \delta_{\mu\alpha\beta}^{\nu\gamma\delta} (-2 \xi_\nu \partial_\gamma h\indices{^\beta_\delta} + \partial_\gamma \xi_\nu h\indices{^\beta_\delta}) \right) + 2 G(h)_{\alpha\beta} \xi^\beta
\end{equation}
which is co-exact on-shell. Therefore, the charges $Q(\alpha)$ associated with such $\alpha$ vanish on-shell. The distinct charges are labelled by tensors $\alpha$ satisfying \eqref{eq:G(alpha)=0} modulo those of the form \eqref{eq:alpha=dxi}.

We demonstrate in Appendix~\ref{app:quantisation} in a canonical quantisation framework that the operator $Q(\alpha)$ satisfies
\begin{equation}
    \comm{h_{ij}(\vec{x})}{Q(\alpha)} = i\alpha_{ij}(\vec{x})
\end{equation}
where $i,j$ denote spatial components and we introduced the equal time commutator.\footnote{In the quantum theory, $h_{00}$ and $h_{0i}$ enter only as Lagrange multipliers and so are not physical fields (see Appendix~\ref{app:quantisation}).}
Here $\Sigma_{d-1}$ is chosen as a region within a constant time hypersurface and we assume that $\vec{x} \equiv (x^1,\dots,x^{d-1}) \in \Sigma_{d-1}$, otherwise the commutator vanishes. Then the finite transformation \eqref{eq:0-form-shift} is given by the action of $e^{i\epsilon Q(\alpha)}$. For constant $\alpha$, the charge can be written in terms of a symmetric tensor charge $Z^{\mu\nu}$ with $Q(\alpha)=\frac{1}{2} \alpha_{\mu\nu}Z^{\mu\nu}$.

Under a linearised diffeomorphism \eqref{eq:linear_diffeo}, $j(\alpha)$ transforms by a total derivative
\begin{equation}
    j(\alpha)_\alpha \to j(\alpha)_\alpha + \partial^\beta \left( 3 \delta^{\nu\gamma\delta}_{\mu\alpha\beta} ( -\alpha\indices{^\mu_\nu} \partial_\gamma \xi_\delta + 2 \partial_\gamma \alpha\indices{^\mu_\nu} \xi_\delta) \right)
\end{equation}
where we have used \eqref{eq:G(alpha)=0}.
This guarantees that the integrated operator \eqref{eq:Q(alpha)} is a well-defined operator in the quantum theory.

While higher-form symmetries are necessarily abelian \cite{Gaiotto2015GeneralizedSymmetries}, the introduction of the 2-tensor $\alpha$ in the construction implies that $Q(\alpha)$ generates a 0-form symmetry. Therefore, its charge algebra can in principle encode a richer structure. This has been seen for $p$-form gauge fields in \cite{Hofman:2018lfz}, where the analogous 0-form symmetry charges obey a centrally-extended algebra reminiscent of Kac-Moody algebras familiar in two dimensions. The representation theory of these algebras was studied recently in \cite{Hofman:2024oze} in the construction of the Hilbert space of four-dimensional Maxwell theory on an arbitrary spatial topology. 
This structure is also present for the charges $Q(\alpha)$ which, as shown in Appendix~\ref{app:quantisation}, satisfy a centrally-extended algebra
\begin{equation}\label{eq:central_ext}
    \comm{Q(\alpha)}{Q(\alpha')} = i\int_\Sigma \star \chi(\alpha,\alpha')
\end{equation}
where $\chi(\alpha,\alpha')$ is a 1-form with components
\begin{equation}\label{eq:zeta_def}
    \chi(\alpha,\alpha')_\mu = 3\alpha\indices{_{[\mu}^\rho} (\dd_L\alpha')\indices{_{\rho\sigma]|}^\sigma} - (\alpha\leftrightarrow\alpha')
\end{equation}
In the simpler case where $\dd_L \alpha = \dd_L \alpha' =0$, the central extension vanishes and the $Q(\alpha)$ are abelian.

We can also define codimension-2 topological operators by contracting $J(h)$ with a constant vector $N^\mu$, to give a conserved 2-form current $j(N)_{\mu\nu} = J_{\mu\nu|\rho}N^\rho$. The associated charge
\begin{equation}\label{eq:Q(N)}
    Q(N) = \int_{\Sigma_{d-2}} \star j(N)
\end{equation}
with $\Sigma_{d-2}$ a $(d-2)$-cycle gives a topological operator generating a 1-form symmetry, which is necessarily abelian. Again, $j(N)$ shifts by a co-exact quantity under linearised diffeomorphisms and so the charge $Q(N)$ is gauge-invariant.

\subsection{Gauging shift symmetries of the graviton}
\label{sec:gauging_graviton_shifts}

We now consider the gauging of the shift symmetries introduced in the previous subsection. Since the Noether current $J(h)$ is a [2,1] bi-form,
we will couple the current to a rank-[2,1] gauge field $\gaugefield$ which transforms as
\begin{equation}\label{eq:A_gauge_transf}
    \gaugefield \to \gaugefield+\dd_L \alpha
\end{equation}
Explicitly, the components $\gaugefield_{\mu\nu|\rho}$ of $\gaugefield$ transform as 
\begin{equation}
     \gaugefield_{\mu\nu|\rho} \to \gaugefield_{\mu\nu|\rho} +  \partial_{[\mu} \alpha_{\nu]\rho}
\end{equation}

The simplest way to gauge the shift symmetry is to note that the combination 
\begin{equation}\label{eq:ggamma}
    \ggamma(h) = \Gamma(h) - \gaugefield
\end{equation}
is invariant under the transformations \eqref{eq:graviton_shift} and \eqref{eq:A_gauge_transf}. Furthermore, inserting \eqref{eq:G(h)} into \eqref{eq:S_FP} and integrating by parts, the graviton action \eqref{eq:S_FP} can be written as
\begin{equation}\label{eq:S_FP_Gamma}
    S_{\text{FP}} = -\frac{3}{2} \int_{\mathcal{M}} \dd[d]{x} \Gamma(h)\indices{_{\mu\rho|}^{[\mu}} \Gamma(h)\indices{^{\rho\alpha]}_{|\alpha}}
\end{equation}
The shift symmetry \eqref{eq:graviton_shift} can then be gauged simply by replacing $\Gamma(h)$ with $\ggamma(h)$ in this form of the graviton action to give
\begin{equation}\label{eq:S_electric_gauged}
    S^{\text{e}}_{\text{FP}} = -\frac{3}{2} \int_{\mathcal{M}} \dd[d]{x} \ggamma(h)\indices{_{\mu\rho|}^{[\mu}} \ggamma(h)\indices{^{\rho\alpha]}_{|\alpha}}
\end{equation}
which is now gauge-invariant under simultaneous transformations \eqref{eq:graviton_shift} and \eqref{eq:A_gauge_transf}.
This is analogous to the replacement of $F$ by $F-B$ in the Maxwell action to gauge the electric 1-form symmetry in eq.~\eqref{eq:Maxwell_electric_gauged}. 

We note that the gauged theory with action \eqref{eq:S_electric_gauged} is invariant under linearised diffeomorphisms provided that $\gaugefield$ transforms as a connection, i.e.\ the action is invariant under
\begin{equation}
\label{Cdiff}
    h_{\mu\nu} \to h_{\mu\nu} + 2\partial_{(\mu}\xi_{\nu)}\qc \gaugefield_{\mu\nu|\rho} \to \gaugefield_{\mu\nu|\rho} + \partial_\rho \partial_{[\mu}\xi_{\nu]}
\end{equation}
as these transformations leave $\ggamma$ invariant. This is of course a special case of the full gauge symmetry \eqref{eq:graviton_shift} and \eqref{eq:A_gauge_transf} for which $\alpha_{\mu\nu} = 2 \partial_{(\mu}\xi_{\nu)}$.
\subsection{Dual symmetry}
\label{sec:dual_symmetries}

There is also a conserved $[d-2,1]$ bi-form current given by
\begin{equation}\label{eq:Jtilde}
    \tilde{J}(h) = \star_L \Gamma(h)
\end{equation}
which satisfies
\begin{equation}\label{eq:Jtilde_conserved}
    \dd^\dag_L \tilde{J}(h) = 0
\end{equation}
as a result of $\dd_L \Gamma(h) = 0$, which follows from \eqref{eq:Gamma}. We note that this current is not right-conserved, $\dd^\dag_R \tilde{J}(h) \neq 0$, and is also not irreducible under $GL(d,\mathbb{R})$. This current is associated with a $[d-3,1]$ bi-form symmetry. 
The current $\tilde{J}(h)$ and its derivatives can be contracted with suitable tensors to give topological operators as discussed earlier for $J(h)$; the topological operators constructed from $\tilde{J}(h)$ will be discussed elsewhere.
The current $\tilde{J}(h)$ is not gauge-invariant and transforms by
\begin{equation}
    \delta \tilde{J}(h)\indices{_{\mu_1\dots\mu_{d-2}|}^\nu} = \frac{1}{2} \epsilon_{\mu_1\dots\mu_{d-2}\alpha\beta} \partial^\nu \partial^\alpha \xi^\beta
\end{equation}
under a linearised diffeomorphism \eqref{eq:linear_diffeo}.

The symmetry associated with the current \eqref{eq:Jtilde} can be gauged by coupling $\tilde{J}(h)$ to a rank-$[d-2,1]$ background tensor gauge field $\gaugefieldtilde$ which transforms as
\begin{equation}\label{eq:Atilde_transf}
    \gaugefieldtilde \to \gaugefieldtilde + \dd_L \tilde{\alpha}
\end{equation}
with $\tilde{\alpha}$ a $[d-3,1]$ parameter.
The gauging is achieved by adding this coupling to the  action \eqref{eq:S_FP_Gamma} to give\footnote{The prefactor of the $\tilde{J}\gaugefieldtilde$ term is chosen for convenience and could be absorbed into a rescaling of $\tilde{J}(h)$.}
\begin{align}
\begin{split}\label{eq:S_magnetic_gauged}
    S_{\text{FP}}^{\text{m}} &= -\frac{3}{2} \int_{\mathcal{M}} \dd[d]{x} \Gamma(h)\indices{_{\mu\rho|}^{[\mu}} \Gamma(h)\indices{^{\rho\alpha]}_{|\alpha}} -\frac{2}{d!} \int_{\mathcal{M}} \dd[d]{x} \tilde{J}(h)_{\mu_1\dots\mu_{d-2}|\nu} \gaugefieldtilde^{\mu_1\dots\mu_{d-2}|\nu} \\
    &= -\frac{3}{2} \int_{\mathcal{M}} \dd[d]{x} \Gamma(h)\indices{_{\mu\rho|}^{[\mu}} \Gamma(h)\indices{^{\rho\alpha]}_{|\alpha}} + \int_{\mathcal{M}} \Gamma(h) \,\wedgedot\, \gaugefieldtilde
\end{split}
\end{align}
which is then invariant under \eqref{eq:Atilde_transf}.
In the last line, we are using the notation introduced in \eqref{eq:wedgedot}.
Note that the gauging of this symmetry does not involve a transformation of the graviton, just as the magnetic $(d-3)$-form symmetry of Maxwell theory did not involve a transformation of the photon $A$.
As in section~\ref{sec:graviton_shifts}, we can construct topological operators from this $[d-2,1]$ bi-form current $\tilde{J}$ by contracting with other tensors which satisfy some differential condition.

\section{The dual graviton theory and its symmetries}
\label{sec:shifts_for_dual}

\subsection{Shift symmetry for the dual graviton}

As we have seen, Maxwell theory can be formulated in terms of the photon $A$ or the dual photon $\tilde A$ and either formulation has a shift symmetry. The current $J_e \sim F$ is the conserved Noether current associated with the shift symmetry of $A$, while $J_m \sim \star F$ is the Noether current for the shift symmetry of $\tilde A$ in the dual formulation.
Linearised gravity can be formulated in terms of the graviton $h$ as above, but also has a dual formulation in terms of a dual graviton $D$ \cite{Hull2000, Hull2001DualityFields}. See \cite{HullYetAppear} for a discussion of the dual theory and further references.
The Noether current for the shift symmetry for the graviton is $J(h)$. In this section, we will investigate the shift symmetry for the dual graviton in the dual formulation and then in the following section we will relate the conserved current for this to $\tilde{J}$.

The dual formulation of a free spin-two field is in terms of the dual graviton, which is an irreducible $[n,1]$ tensor gauge field $D$, where $n\equiv d-3$. Its gauge transformations are
\begin{equation}\label{Dgauge}
    D\to D+ \dd_L \zeta + \mathcal{Y}_{[n,1]} \left(\dd_R \phi\right)
\end{equation}
where $\zeta$ is an $[n-1,1]$ bi-form and $\phi$ is an $n$-form. Explicitly, 
\begin{equation}
    D_{\mu_1\dots\mu_n|\nu} \to D_{\mu_1\dots\mu_n|\nu} + \partial_{[\mu_1} \zeta_{\mu_2\dots\mu_n]|\nu} + \partial_\nu \phi_{\mu_1\dots\mu_n} -  \partial_{[\nu} \phi_{\mu_1\dots\mu_n]}
\end{equation}
There are two distinct connections which can be constructed from $D$ \cite{deMedeiros:2003osq}. The first is an irreducible $[n+1,1]$ bi-form
\begin{equation}\label{eq:Gammatilde}
    \tilde{\Gamma}(D) = \dd_L D
\end{equation}
while the second is a reducible $[n,2]$ bi-form
\begin{equation}
    \hat{\Gamma}(D) = \dd_R D
\end{equation}
We note that, while $\tilde{\Gamma}$ transforms in an irreducible representation of $GL(d,\mathbb{R})$, $\hat{\Gamma}$ does not. Therefore, while it has the index symmetries $\hat{\Gamma}(D)_{\mu_1\dots\mu_n|\nu_1\nu_2} = \hat{\Gamma}(D)_{[\mu_1\dots\mu_n]|[\nu_1\nu_2]}$, it is not true that $\hat{\Gamma}(D)_{[\mu_1\dots\mu_n|\nu_1]\nu_2}=0$ in general.

The gauge-invariant curvature is an irreducible $[n+1,2]$ bi-form which can be written
\begin{equation}
    S(D) = \dd_R \tilde{\Gamma}(D) = \dd_L \hat{\Gamma}(D)
\end{equation}
The action \cite{Curtright:1980yk} for a tensor gauge field in this representation can be written as \cite{Medeiros2003ExoticDuality}
\begin{equation}\label{eq:S_dual}
    S_{\text{dual}} = \frac{1}{2} \int_{\M} \dd[d]{x} D_{\mu_1\dots\mu_n|\nu} E(D)^{\mu_1\dots\mu_n|\nu}
\end{equation}
where 
\begin{equation}
    E(D)\indices{_{\mu_1\dots\mu_n|}^\nu} = \delta_{\mu_1\dots\mu_n\alpha\beta}^{\rho_1\dots\rho_{n+1}\nu} S(D)\indices{_{\rho_1\dots\rho_{n+1}|}^{\alpha\beta}}
\end{equation}
is the Einstein tensor for the dual graviton, which is an irreducible $[n,1]$ bi-form. Varying the action with respect to $D$ gives the free field equation
\begin{equation}
    E(D)_{\mu_1\dots\mu_n|\nu} = 0
\end{equation}

We now study the shift symmetry of the dual graviton theory.
Consider the transformation
\begin{equation}\label{eq:D_shift}
    D_{\mu_1\dots\mu_n|\nu} \to D_{\mu_1\dots\mu_n|\nu} + \tilde{\alpha}_{\mu_1\dots\mu_n|\nu}
\end{equation}
where $\tilde{\alpha}$ is an irreducible $[n,1]$ bi-form. Under this variation, the action \eqref{eq:S_dual} changes by
\begin{equation}
    \delta S_{\text{dual}} = \int_{\mathcal{M}} \dd[d]{x} D_{\mu_1\dots\mu_n|\nu} E(\tilde{\alpha})^{\mu_1\dots\mu_n|\nu}
\end{equation}
so that \eqref{eq:D_shift} is a global symmetry provided that $E(\tilde{\alpha})=0$.\footnote{The action \eqref{eq:S_dual} is the most relevant term in the effective theory of a $[n,1]$-tensor gauge field on Minkowski space which could, in general, have higher-derivative terms involving the curvature $S(D)$. The shift \eqref{eq:D_shift} would then  remain a symmetry provided that we impose the stronger constraint $S(\tilde{\alpha}) = \dd_L \dd_R \tilde{\alpha} = 0$. If we consider only the leading term \eqref{eq:S_dual} in the EFT expansion, we need only require that $E(\tilde{\alpha})=0$.}
The variation of the action under a transformation with an unconstrained parameter $\tilde{\alpha}$ can be written as $\dd_L\tilde \alpha$ contracted with an irreducible $[n+1,1]$ bi-form $\tilde{J}(D)$
which is given by
\begin{equation}\label{eq:Jtilde_D}
    \tilde{J}(D)\indices{_{\mu_1\dots\mu_{n+1}|}^\nu} = \frac{n+1}{2} \delta^\nu_{[\beta} \tilde{\Gamma}(D)\indices{_{\mu_1\dots\mu_{n+1}]|}^\beta}
\end{equation}
with a normalisation chosen for later convenience.
This is then a Noether current for the $\tilde \alpha$ global symmetry.
(As usual, this is defined only up to the addition of a co-exact term of the form $\dd_L^\dagger \tilde{H}$ for some $[n+2,1]$ bi-form $\tilde{H}$.)
We then find that $\dd_L^\dag \tilde{J}(D) \propto E(D)$ and so on-shell
(i.e.\ when $E(D)=0$) $\tilde{J}(D)$ satisfies the left-conservation law
$\dd_L^\dag \tilde{J}(D)=0$.

There is a gauge-invariant codimension-1 topological operator $\tilde{Q}(\tilde{\alpha})$ which generates a shift of $D$ by a given $\tilde{\alpha}$ in the quantum theory. The construction is analogous to the one in section~\ref{sec:graviton_shifts} and the $\tilde{Q}(\tilde{\alpha})$ also satisfy a centrally-extended algebra. For example, in $d=4$, this is the same algebra as for the graviton theory found in \eqref{eq:central_ext}.

The shift symmetry \eqref{eq:D_shift} of the dual graviton theory can be gauged by introducing a background (irreducible) $[n+1,1]$ tensor gauge field $\gaugefieldtilde$ which transforms as in eq.~\eqref{eq:Atilde_transf}.
The combination 
\begin{equation}
    \tilde{\ggamma}(D) = \tilde{\Gamma}(D) - \gaugefieldtilde
\end{equation}
is invariant under the combined transformations \eqref{eq:D_shift} and \eqref{eq:Atilde_transf}. We note that the action \eqref{eq:S_dual} can be written as
\begin{equation}\label{eq:S_dual_rewrite_tilde}
    S_{\text{dual}} = -\frac{1}{2} \int_{\mathcal{M}} \dd[d]{x} \tilde{\Gamma}(D)\indices{^{\mu_1\dots\mu_{n+1}|}_{[\mu_1}} \tilde{\Gamma}(D)\indices{_{\mu_2\dots\mu_{n+1}\alpha]|}^\alpha}
\end{equation}
by integrating by parts. Hence, the shift symmetry can be gauged by replacing $\tilde{\Gamma}(D)$ with $\tilde{\ggamma}(D)$, giving
\begin{equation}
    S^{\text{e}}_{\text{dual}} = -\frac{1}{2} \int_{\mathcal{M}} \dd[d]{x} \tilde{\ggamma}(D)\indices{^{\mu_1\dots\mu_{n+1}|}_{[\mu_1}} \tilde{\ggamma}(D)\indices{_{\mu_2\dots\mu_{n+1}\alpha]|}^\alpha}
\end{equation}
This action is also invariant under the gauge transformations for $D$ in \eqref{Dgauge} provided the background gauge field $\gaugefieldtilde$ transforms as 
\begin{equation}\label{Ctgauge}
    \gaugefieldtilde\to \gaugefieldtilde + \dd_L  \mathcal{Y}_{[n,1]} \left(\dd_R \phi\right)
\end{equation}

Alternatively, the variation of the action under a transformation with an unconstrained parameter $\tilde{\alpha}$ can be written as $\dd_R\tilde{\alpha}$ contracted with a (reducible) $[n,2]$ bi-form $\hat{J}(D)$ which is then a further Noether current associated with the $\tilde{\alpha}$ global symmetry and is given by 
\begin{equation}\label{eq:Jhat}
    \hat{J}(D)\indices{_{\mu_1\dots\mu_n|}^{\nu_1\nu_2}} = \delta^{\nu_1\nu_2}_{[\alpha\beta} \hat{\Gamma}(D)\indices{_{\mu_1\dots\mu_n]|}^{\alpha\beta}}
\end{equation}
This satisfies $\dd_R^\dag \hat{J}(D) \propto E(D)$ and so 
obeys the right-conservation law $\dd_R^\dag \hat{J}(D)=0$ on-shell.

Taking $\hat{J}(D)$ in eq.~\eqref{eq:Jhat} as the Noether current provides another way to gauge the shift symmetry of the dual graviton. The action \eqref{eq:S_dual} can be rewritten as
\begin{equation}\label{eq:S_dual_rewrite_hat}
    S_{\text{dual}} = \frac{(-1)^n}{2} \int_{\mathcal{M}} \dd[d]{x} \hat{\Gamma}(D)\indices{^{\mu_1\dots\mu_n}_{|[\mu_1\mu_2}} \hat{\Gamma}(D)\indices{_{\mu_3\dots\mu_n \alpha\beta]|}^{\alpha\beta}}
\end{equation}
Now, the combination
\begin{equation}
    \hat{\ggamma}(D) = \hat{\Gamma}(D) - \gaugefieldhat
\end{equation}
is invariant under the shift \eqref{eq:D_shift} where $\gaugefieldhat$ is a background (reducible) $[n,2]$ tensor gauge field which transforms as
\begin{equation}
    \gaugefieldhat \to \gaugefieldhat + \dd_R \tilde{\alpha}
\end{equation}
under the symmetry.\footnote{$\gaugefieldhat$ is in the same $GL(d,\mathbb{R})$ representation as $\hat{\ggamma}$; that is, it has index symmetries $\gaugefieldhat_{\mu_1\dots\mu_n|\nu_1\nu_2} = \gaugefieldhat_{[\mu_1\dots\mu_n]|[\nu_1\nu_2]}$ but is not irreducible.} Then the gauging is simply achieved by replacing $\hat{\Gamma}(D)$ with $\hat{\ggamma}(D)$ in eq.~\eqref{eq:S_dual_rewrite_hat}, giving
\begin{equation}
    S_{\text{dual}}^{\text{e}\,\prime} = \frac{(-1)^n}{2} \int_{\mathcal{M}} \dd[d]{x} \hat{\ggamma}(D)\indices{^{\mu_1\dots\mu_n}_{|[\mu_1\mu_2}} \hat{\ggamma}(D)\indices{_{\mu_3\dots\mu_n \alpha\beta]|}^{\alpha\beta}}
\end{equation}
This action is invariant under the gauge transformations \eqref{Dgauge} provided the gauge field transforms as 
\begin{equation}
    \gaugefieldhat\to \gaugefieldhat + \dd_R  \left( \dd_L \zeta + \mathcal{Y}_{[n,1]} \left(\dd_R \phi\right) \right)
\end{equation}

\subsection{Dual symmetry for the dual graviton}

The graviton theory has an identically conserved current \eqref{eq:Jtilde} which was coupled to a gauge field $\gaugefieldtilde$ in section~\ref{sec:dual_symmetries}.
In the same way, the dual graviton theory has
a conserved $[2,1]$ bi-form current
\begin{equation}\label{eq:JD}
     {J}(D) = \star_L \tilde \Gamma(D)
\end{equation}
which satisfies
\begin{equation}
    \dd^\dag_L J(D) = 0
\end{equation}
as a result of $\dd_L \tilde\Gamma(D) = 0$.

This Noether current can be coupled to a background irreducible [2,1] tensor gauge field $\gaugefield$ which transforms as in \eqref{eq:A_gauge_transf}. The gauging is implemented by coupling $J$ to the gauge field,
\begin{align}
\begin{split}
    S_{\text{dual}}^{\text{m}} &= -\frac{1}{2} \int_{\mathcal{M}} \dd[d]{x} \tilde{\Gamma}(D)\indices{^{\mu_1\dots\mu_{n+1}|}_{[\mu_1}} \tilde{\Gamma}(D)\indices{_{\mu_2\dots\mu_{n+1}\alpha]|}^\alpha} - \frac{1}{d(d-1)} \int_{\mathcal{M}} \dd[d]{x} J(D)_{\mu\nu|\rho}\gaugefield^{\mu\nu|\rho} \\
    &= -\frac{1}{2} \int_{\mathcal{M}} \dd[d]{x} \tilde{\Gamma}(D)\indices{^{\mu_1\dots\mu_{n+1}|}_{[\mu_1}} \tilde{\Gamma}(D)\indices{_{\mu_2\dots\mu_{n+1}\alpha]|}^\alpha} + \int_{\mathcal{M}} \tilde{\Gamma}(D) \,\wedgedot\, \gaugefield
\end{split}
\end{align}
This is invariant under the gauge transformation \eqref{eq:A_gauge_transf} with $D$ invariant.

\section{Relation between the graviton and dual graviton theories}
\label{sec:relating_dual_theories}

For Maxwell theory, the field strength $F$ can be written locally in terms of the potential $A$ as $F=F(A)$ with $F(A)= \dd A$ or in terms of the dual potential $\tilde A$ as $F=\star \tilde F(\tilde A)$ where $\tilde F(\tilde A)= \dd\tilde A$. The two formulations are then related by
\begin{equation}
     \tilde F(\tilde A) = \star F(A)
\end{equation}
The situation is similar for linearised gravity. The $[2,2]$ bi-form field strength $R$ can be written as $R=R(h)$ where $R(h)=-2\dd_L \dd_R h$ or as $R=\star_L S(D)$ where $S(D)=\dd_L \dd_R D$. This then gives the duality relation \cite{Hull2001DualityFields}
\begin{equation}
    S(D) = \star_L R(h)
\end{equation}

We can fix a gauge where not only the curvatures are related in this way, but the connections $\Gamma(h)$ and $\tilde{\Gamma}(D)$ are also dual \cite{HullYetAppear}. In terms of the graviton description, the relevant gauge is one in which
\begin{equation}\label{eq:h_duality_gauge}
    \Gamma(h)\indices{_{\mu\nu|}^\nu} = 0
\end{equation}
while in the dual graviton description we must impose
\begin{equation}\label{eq:D_duality_gauge}
    \tilde{\Gamma}(D)\indices{_{\mu_1\dots\mu_n\nu|}^\nu} = 0
\end{equation}
With these gauge choices, we can consistently impose the duality at the level of the connections:
\begin{equation}\label{eq:connection_duality}
    \Gamma(h) = -\frac{1}{2}\star_L \tilde{\Gamma}(D)
\end{equation}
(Note there is no such relation for $\hat{\Gamma}(D)$.)

With the gauge choice \eqref{eq:D_duality_gauge}, the current $\tilde{J}(D)$ in \eqref{eq:Jtilde_D} simply becomes
\begin{equation}
    \tilde{J}(D) = \frac{1}{2} \tilde{\Gamma}(D)
\end{equation}
which is the same current as was introduced in eq.~\eqref{eq:Jtilde} once we impose the duality \eqref{eq:connection_duality}, i.e.
\begin{equation}
    \tilde{J}(D) = \tilde{J}(h)
\end{equation}
Therefore, the current $\tilde{J}(h)$ in the original graviton formulation of the theory can be interpreted as coming from a shift symmetry in the dual field description. This is precisely the same structure as was seen in Maxwell theory in section~\ref{sec:maxwell}.

We have seen that the shift symmetry of the dual graviton \eqref{eq:D_shift} is related to the identically conserved current \eqref{eq:Jtilde} in the original graviton theory via duality.
Similarly, the current $J(h)$ associated with the shift symmetry of the graviton \eqref{eq:graviton_shift} is an identically conserved current in the dual theory. Explicitly, in the gauge \eqref{eq:h_duality_gauge}, we find $J(h) = \Gamma(h)$ and so $J$ can be written in terms of the dual graviton as
\begin{equation}
    J(h) = \Gamma(h) = - \frac{1}{2} \star_L \tilde{\Gamma}(D) \equiv J(D)
\end{equation}
where we have used the duality \eqref{eq:connection_duality} in the second equality. 

\section{Mixed 't Hooft anomaly}
\label{sec:anomalies}

In sections~\ref{sec:graviton_shifts} and \ref{sec:dual_symmetries} we studied a pair of dual symmetries of the free graviton theory and gave their gauging in eqs.~\eqref{eq:S_electric_gauged} and \eqref{eq:S_magnetic_gauged}, respectively. This required the introduction of background tensor gauge fields $\gaugefield$ and $\gaugefieldtilde$. Let us now introduce both background fields and seek to simultaneously gauge the two symmetries:
\begin{equation}\label{eq:S_both_gauged}
    S_{\text{FP}}^{\text{e}+\text{m}} = -\frac{3}{2} \int_{\mathcal{M}} \dd[d]{x} \ggamma(h)\indices{_{\mu\rho|}^{[\mu}} \ggamma(h)\indices{^{\rho\alpha]}_{|\alpha}} + \int_{\mathcal{M}} \Gamma(h) \,\wedgedot\, \gaugefieldtilde
\end{equation}
where $\ggamma(h) \equiv \Gamma(h) - \gaugefield$ as before. We recall the field transformations here for convenience:
\begin{gather}
    h \to h + \alpha \qc \gaugefield \to \gaugefield + \dd_L \alpha \label{eq:electric_transf_repeat} \\
    \gaugefieldtilde \to \gaugefieldtilde + \dd_L \tilde{\alpha} \label{eq:magnetic_transf_repeat}
\end{gather}
The action \eqref{eq:S_both_gauged} is invariant under the transformation \eqref{eq:magnetic_transf_repeat}. However, while the first term in \eqref{eq:S_both_gauged} is invariant under the transformation \eqref{eq:electric_transf_repeat}, the second term is not and varies by
\begin{equation}\label{eq:anomaly}
    \delta S_{\text{FP}}^{\text{e}+\text{m}} = \int_{\mathcal{M}} \dd_L \alpha \,\wedgedot\, \gaugefieldtilde
\end{equation}
This variation can be cancelled by adding a counterterm $-\frac{1}{2} \gaugefield \,\wedgedot\, \gaugefieldtilde$ so that the action becomes
\begin{equation}\label{eq:S_both_gaugedplus}
     \hat S_{\text{FP}}^{\text{e}+\text{m}} = -\frac{3}{2} \int_{\mathcal{M}} \dd[d]{x} \ggamma(h)\indices{_{\mu\rho|}^{[\mu}} \ggamma(h)\indices{^{\rho\alpha]}_{|\alpha}} + \int_{\mathcal{M}} \ggamma(h) \,\wedgedot\, \gaugefieldtilde
\end{equation}
This replaces $\Gamma(h)$ by $\ggamma(h)$ in the second term of eq.~\eqref{eq:S_both_gauged} so that it is invariant under \eqref{eq:electric_transf_repeat}. However, adding this term breaks the magnetic shift symmetry and the resulting action is no longer invariant under \eqref{eq:magnetic_transf_repeat}, transforming as
\begin{equation}\label{eq:anomalycc}
    \delta \hat S_{\text{FP}}^{\text{e}+\text{m}} = -\int_{\mathcal{M}} \dd_L \tilde\alpha \,\wedgedot\, {\gaugefield}
\end{equation}
There is no counterterm involving the background fields $C$ and $\tilde{C}$ which can be added to make the action invariant under both \eqref{eq:electric_transf_repeat} and \eqref{eq:magnetic_transf_repeat}.
The impossibility of consistently coupling background gauge fields for both symmetries signals a mixed 't Hooft anomaly between the $[1,1]$ bi-form symmetry associated with the shift \eqref{eq:graviton_shift} and the $[d-3,1]$ bi-form symmetry associated with the dual shift \eqref{eq:D_shift}.
The anomaly \eqref{eq:anomaly} is analogous to eq.~\eqref{eq:Maxwell_anomaly} in Maxwell theory. The gauge-invariant $(d+2)$-form anomaly polynomial which leads to to eq.~\eqref{eq:anomaly} by descent is\footnote{We note that the contraction of indices involved in the $\wedgedot$ product implies that the anomaly inflow Lagrangian $\gaugefield \,\wedgedot\, \dd_L \gaugefieldtilde$ depends explicitly on the background Minkowski metric and that unavoidable metric dependence means that the $(d+1)$-dimensional field theory is not properly topological.} 
\begin{equation}\label{eq:anomaly_poly}
    \mathcal{I}_{d+2} = \dd_L \gaugefield \,\wedgedot\, \dd_L \gaugefieldtilde
\end{equation}

The second term in the action \eqref{eq:S_both_gauged} is not invariant under the gauge transformations \eqref{Cdiff} but the variation is cancelled by adding the counterterm to give the action \eqref{eq:S_both_gaugedplus} which is invariant under these transformations provided that $\gaugefieldtilde$ is invariant.
Thus cancelling the anomaly in the transformations \eqref{eq:magnetic_transf_repeat} also cancels the anomaly in the gauge transformations \eqref{Cdiff}, as was to be expected, as they are a special case of the transformations \eqref{eq:magnetic_transf_repeat}.

In the dual theory, formulated in terms of $D$, the anomaly can be uncovered by simultaneously coupling the currents for both symmetries to background gauge fields using the couplings in section~\ref{sec:shifts_for_dual}, i.e.
\begin{equation}
    S_{\text{dual}}^{\text{e}+\text{m}} = -\frac{1}{2} \int_{\mathcal{M}} \dd[d]{x} \tilde{\ggamma}(D)\indices{^{\mu_1\dots\mu_{n+1}|}_{[\mu_1}} \tilde{\ggamma}(D)\indices{_{\mu_2\dots\mu_{n+1}\alpha]|}^\alpha} + \int_{\mathcal{M}} \tilde{\Gamma}(D) \,\wedgedot\, \gaugefield
\end{equation}
This is invariant under \eqref{eq:A_gauge_transf} but under the combined transformations \eqref{eq:D_shift} and \eqref{eq:Atilde_transf} it transforms as
\begin{equation}\label{eq:anomalyd}
    \delta S_{\text{dual}}^{\text{e}+\text{m}} = \int_{\mathcal{M}} \dd_L \tilde\alpha \,\wedgedot\, {\gaugefield}
\end{equation}
in agreement with \eqref{eq:anomalycc} (up to a factor which could have been absorbed into a redefinition of the currents). The anomaly is then seen again in the dual theory.

\section{Implications of the anomaly}
\label{sec:massless_graviton}

\subsection{The currents}

The anomaly of the last section can be understood in terms of the associated currents $J$ and $\tilde{J}$. 
These currents are not invariant under linearised diffeomorphisms \eqref{eq:linear_diffeo}, but, from \eqref{eq:J(h)_conservations}, $J$ is left-conserved on-shell and, from \eqref{eq:Jtilde_conserved}, $\tilde{J}$ is left-conserved identically.
Furthermore, from \eqref{eq:J_right_conserved}, $J$ is right-conserved identically.

From these currents, we can build tensors which are gauge-invariant under \eqref{eq:linear_diffeo}. Namely, the [2,2] curvature bi-form $R$ defined in \eqref{eq:riemann} is left-conserved on-shell,
\begin{equation}
    \dd_L^\dag R \doteq 0
\end{equation}
as a result of the on-shell left-conservation of $J$ (see \eqref{eq:X_def}). Also, we can build the left-dual curvature from $\tilde{J}$ by
\begin{equation}\label{eq:Y_def}
    \star_L R(h) = -2\dd_R \tilde{J}(h)
\end{equation}
which is gauge-invariant and left-conserved,
\begin{equation}
    \dd_L^\dag (\star_L R) = 0
\end{equation}

Let us now consider the conserved currents in the gauged theory \eqref{eq:S_electric_gauged} where we have coupled a [2,1] background field $C$ such that the shift \eqref{eq:graviton_shift} is a symmetry for arbitrary $\alpha_{\mu\nu}(x)$.
In the gauged theory \eqref{eq:S_electric_gauged}, the equations of motion are no longer $G_{\mu\nu}=0$ but rather $\mathbf{G}_{\mu\nu}=0$ where $\mathbf{G}_{\mu\nu}$ is the Einstein tensor constructed from $\ggamma$ (defined in \eqref{eq:ggamma}) rather than $\Gamma$.
Therefore, those quantities which were conserved on-shell in the ungauged theory will, in general, not be conserved on-shell in the gauged theory. However, in the gauged theory, we can build a [2,1] current
\begin{equation}
    \mathbf{J}\indices{_{\mu\nu|}^\rho} = -3 \delta_{\mu\nu\alpha}^{\sigma\rho\beta} \ggamma\indices{_{\sigma\beta|}^\alpha}
\end{equation}
which is left-conserved on-shell,
\begin{equation}\label{eq:improved_J}
    \dd_L^\dag \mathbf{J} \doteq 0
\end{equation}
This `improved' current $\mathbf{J}$ is the natural modification of $J(h)$ in \eqref{JGam} such that it is invariant under the (now gauged) shift symmetry \eqref{eq:graviton_shift}. 
Similarly, the dual current
\begin{equation}
    \tilde{\mathbf{J}} = \star_L \ggamma
\end{equation}
is the improvement of \eqref{eq:Jtilde} which is invariant under the gauge symmetry \eqref{eq:graviton_shift}. 
However, $\tilde{\mathbf{J}}$ is \emph{not} left-conserved in the gauge theory if the background field configuration $C$ is not flat (that is, if $\dd_L C \neq0$), since
\begin{equation}\label{eq:improved_Jtilde}
    \dd_L^\dag \tilde{\mathbf{J}} = 3 (-1)^{d-1} \star_L \dd_L C
\end{equation}
This is the analogue of the statement that $F-B$ is not closed in Maxwell theory once the electric 1-form symmetry is gauged, as seen at the end of section~\ref{sec:maxwell}. This is a manifestation of the mixed 't Hooft anomaly: gauging the graviton shift \eqref{eq:graviton_shift} explicitly breaks the dual symmetry.

Similarly, there are improved versions of the [2,2] current $R$ and the $[d-2,2]$ current $\star_L R$ in the gauged theory. These are given by replacing $\Gamma$ with $\mathbf{\Gamma}$ in the definition of $R$, i.e.\
\begin{equation}
    \mathbf{R} = -2 \dd_R \mathbf{\Gamma}
\end{equation}
Again, it follows from \eqref{eq:improved_J} that $\mathbf{R}$ is left-conserved on-shell,
\begin{equation}\label{eq:XY_anomaly1}
    \dd_L^\dag \mathbf{R} = 0
\end{equation}
and it follows from \eqref{eq:improved_Jtilde} that $\star_L \mathbf{R}$ is \emph{not} left-conserved in the gauged theory. Instead, it satisfies
\begin{equation}\label{eq:XY_anomaly2}
    \dd_L^\dag (\star_L \mathbf{R}) = 3(-1)^{d-1} \star_L \dd_L \dd_R C
\end{equation}
and so is not left-conserved in the presence of a non-flat background $C$ for the graviton shift symmetry.
Moreover, $\mathbf{R}$ fails to be traceless on-shell,
\begin{equation}\label{eq:trace_anom}
    \mathbf{R}\indices{^\mu_{\nu\mu\sigma}} \doteq \partial^\mu C_{\mu\nu|\sigma} + \partial_\sigma C\indices{_{\nu\mu|}^\mu}
\end{equation}
The fact that the improved quantities are not conserved or traceless in the presence of the background gauge field is a signal of the mixed 't Hooft anomaly.

Instead of coupling the [1,1] shift symmetry to a background [2,1] gauge field $C$, we could have coupled the $[d-3,1]$ bi-form symmetry to a background $[d-2,1]$ gauge field $\tilde{C}$ as in \eqref{eq:S_magnetic_gauged}. This can be achieved by adding counterterms to the action involving only the background fields $C$ and $\tilde{C}$.
The equation of motion in this gauged theory is then
\begin{equation}
    G_{\mu\nu} = \frac{1}{d} (\star_L \dd_L \tilde{C})_{(\mu|\nu)}
\end{equation}
We find that $\dd_L^\dag \tilde{J} = 0$ still holds in the gauged theory, but $\dd_L^\dag J \neq0$ when the $\tilde{C}$ background field is not flat. In terms of the gauge-invariant bi-forms $R$ and $\star_L R$, this implies that $\dd_L^\dag (\star_L R)=0$ but $\dd_L^\dag R \neq0$ when $\tilde{C}$ is not flat. Again, we see that gauging one symmetry explicitly breaks the other.

\subsection{Currents and correlators}

In the free graviton theory, it is straightforward to calculate the current correlators $\ev*{J \tilde{J}}$ or $\ev*{RR}$ and to check that they have the form implied by the anomaly. We fix the de Donder gauge
\begin{equation}
    \partial^\mu h_{\mu\nu} - \frac{1}{2}\partial_\nu h\indices{^\mu_\mu} = 0
\end{equation}
where the momentum space propagator\footnote{All correlation functions are time-ordered.}
\begin{equation}
    \mathcal{G}_{\mu\nu\alpha\beta}(p) \equiv \int \dd[d]{x} e^{ipx} \ev*{h_{\mu\nu}(x) h_{\alpha\beta}(0)} 
\end{equation}
can be written as \cite{Hinterbichler:2011tt}
\begin{equation}
    \mathcal{G}_{\mu\nu\alpha\beta}(p) = \frac{-i}{p^2} \left( \frac{1}{2} \eta_{\mu\alpha}\eta_{\nu\beta} + \frac{1}{2} \eta_{\mu\beta} \eta_{\nu\alpha} - \frac{1}{d-2} \eta_{\mu\nu} \eta_{\alpha\beta} \right)
\end{equation}
The current-current correlator in this gauge is
\begin{equation}\label{eq:JJ_correlator}
    \int \dd[d]{x} e^{ipx} \ev*{J_{\mu\nu|\rho}(x) (\star_L \tilde{J})^{\alpha\beta|\gamma}(0)} = \frac{i}{p^2} \left( \frac{1}{2} p_{[\mu} p^{[\alpha} \delta^{\beta]}_{\nu]} \delta^\gamma_\rho + \frac{1}{2} p_{[\mu} \delta_{\nu]}^\gamma p^{[\alpha} \delta^{\beta]}_\rho - \frac{1}{2} \eta_{\rho[\mu} p^{[\alpha} \delta^{\beta]}_{\nu]} p^\gamma \right)
\end{equation}
This correlator implies the following correlator involving the gauge invariant [2,2] current $R$:
\begin{equation}
\begin{split}\label{eq:XY_correlator}
    \int \dd[d]{x} e^{ipx} \ev*{R_{\mu\nu|\rho\sigma}(x) R^{\alpha\beta|\gamma\delta}(0)} &= -\frac{4i}{p^2} \bigg( \frac{1}{2} p_{[\mu} p^{[\alpha} \delta^{\beta]}_{\nu]} p_{[\rho} p^{[\gamma} \delta^{\delta]}_{\sigma]} + (\alpha\beta\leftrightarrow\gamma\delta) \\
    &\qquad\qquad - \frac{1}{d-2} p_{[\mu} \eta_{\nu][\rho} p_{\sigma]} p^{[\alpha} \eta^{\beta][\gamma}p^{\delta]} \bigg)
\end{split}
\end{equation}
which exhibits a $p^{-2}$ pole as a consequence of the masslessness of the graviton.
The linearised Riemann tensor is traceless on-shell so that on-shell the linearised curvature becomes the linearised Weyl tensor. In the quantum theory, taking a trace of the $\ev*{RR}$ correlator, we find
\begin{equation}\label{eq:Ricci_correlator}
    \int \dd[d]{x} e^{ipx} \ev*{\eta^{\mu\rho}R_{\mu\nu|\rho\sigma}(x) R^{\alpha\beta|\gamma\delta}(0)} = -i \left( p^{[\alpha} \delta^{\beta]}_{(\nu} p^{[\gamma} \delta^{\delta]}_{\sigma)} + \frac{1}{d-2} \eta_{\nu\sigma} p^{[\alpha} \eta^{\beta][\gamma}p^{\delta]} \right)
\end{equation}
The terms on the right-hand side are polynomial in $p^\mu$. In position space this is a contact term, so the Ricci tensor may not vanish inside a correlator with other insertions of $R$ at the same point in the quantum theory. Away from other insertions, the Ricci tensor vanishes inside any correlation function.

The anomaly is reflected in correlators via Ward identities. It is straightforward to show that 
\begin{equation}
    \int\dd[d]{x} e^{ipx} \ev*{ R_{\mu\nu|\rho\sigma}(x) \partial^{[\lambda} R^{\alpha\beta]|\gamma\delta}(0)} = 0
\end{equation}
so that $\star_L R$ is left-conserved (equivalently, $R$ is left-closed), whereas
\begin{equation}
\begin{split}\label{eq:divX_contact}
    \int\dd[d]{x} e^{ipx} \ev*{ \partial^{\mu} R_{\mu\nu|\rho\sigma}(x) R^{\alpha\beta|\gamma\delta}(0)} &= -\left(p^{[\alpha} \delta^{\beta]}_\nu p_{[\rho} p^{[\gamma} \delta^{\delta]}_{\sigma]} + (\alpha\beta \leftrightarrow\gamma\delta) \right) \\
    & \quad + \frac{2}{d-2} \eta_{\nu[\rho} p_{\sigma]} p^{[\alpha} \eta^{\beta][\gamma} p^{\delta]} 
\end{split}
\end{equation}
so $R$ is not left-conserved in general in the quantum theory. We note that the right-hand side of \eqref{eq:divX_contact} is polynomial in the momenta $p^\mu$ and so, as above, in position space this is a contact term. This equation then takes the standard form of a Ward identity in position space. Therefore, in position space, we find that in this correlator $\star_L R$ is left-conserved everywhere, whereas $R$ is only left-conserved away from the other insertions of $R$. That is, the classical left-conservation of both $R$ and $\star_L R$ is incompatible in the quantum theory and one of them fails to be conserved when inserted at the same spacetime point as the other.

In this presentation, it is the dual current $\star_L R$ which is conserved everywhere and the anomaly is seen as the lack of conservation of $R$. As explained in the previous subsection, we can shift the anomaly between $R$ and $\star_L R$ by adding counterterms involving only the background gauge fields. At the level of the correlation functions, this has the effect of changing the contact terms in the $\ev*{RR}$ correlator such that $R$ is conserved everywhere and $\star_L R$ fails to be conserved when inserted at the same point as the other insertion in the correlator. However, the $p^{-2}$ term in the correlator is not affected by the addition of such counterterms and is fixed by the anomaly.

Since the currents $J$ and $\tilde{J}$ can be understood as the Noether currents associated with shifts of the graviton and its dual, it is then natural to see the graviton as a Nambu-Goldstone mode. More precisely, the graviton could be interpreted as the Nambu-Goldstone mode associated with the [1,1] bi-form shift symmetry \eqref{eq:graviton_shift} and the dual graviton could be interpreted as the Nambu-Goldstone mode associated with the $[d-3,1]$ bi-form shift symmetry \eqref{eq:D_shift}.
This is the gravitational analogue of viewing the photon as the Nambu-Goldstone mode for a 1-form shift symmetry \cite{Gaiotto2015GeneralizedSymmetries}.

\subsection{Implications of the 't Hooft anomaly in theories with bi-form currents}

We have seen that the free graviton theory, which describes a massless spin-2 particle, has bi-form currents $J$ and $\tilde{J}$ with correlator given by \eqref{eq:JJ_correlator} (modulo gauge transformations) which implies that the gauge-invariant currents $R$, $\star_L R$ have correlator \eqref{eq:XY_correlator}.
These correlation functions express the 't Hooft anomaly. 
We now \emph{reverse} our reasoning to consider a theory (e.g.\ a condensed matter system or a field theory) with a [2,1] bi-form current $J$ and $[d-2,1]$ bi-form current $\tilde{J}$ which are left-conserved and with a mixed 't Hooft anomaly described by the anomaly polynomial \eqref{eq:anomaly_poly}. Given the RG invariance of 't Hooft anomalies, we can use the anomaly to learn about the IR dynamics of such a theory.

In particular, arguments of this type have been used in~\cite{Delacretaz:2019brr, Hinterbichler2023GravitySymmetries, Hinterbichler:2024cxn} to demonstrate the presence of a gapless mode in the IR spectrum of theories with certain mixed 't Hooft anomalies. In those works, anomalous conservation equations analogous to \eqref{eq:improved_J} and \eqref{eq:improved_Jtilde} were shown to be sufficient to fix the correlator of the currents, revealing a $p^{-2}$ pole which corresponds to a massless mode. 
The results of~\cite{Weinberg:2020nsn, Distler:2020fzr} then imply that this massless mode has spin-two. These results do not rely on the specifics of the theory, and will apply to any theory with such a mixed 't Hooft anomaly.

There is a subtlety in applying this argument to the anomaly in \eqref{eq:improved_Jtilde} since the currents $J$ and $\tilde{J}$ are not gauge-invariant. 
We can, however, consider a general theory which has conserved currents which are [2,2] and $[d-2,2]$ bi-forms $X$ and $Y$ that are observable, with $X$ satisfying the same conservation conditions as $R$ and $Y$ satisfying the same conservation conditions as $\star_L R$:
\begin{equation}\label{eq:XY_conservation}
    \dd_L^\dag X \doteq 0\qc \dd_L^\dag Y = 0\qc \dd_R^\dag X \doteq 0\qc \dd_R^\dag Y \doteq 0
\end{equation}
and where $X$ is traceless on-shell.
We suppose that the global symmetries generated by $X$ and $Y$ have a mixed 't Hooft anomaly of the same form as that considered above, for example in \eqref{eq:XY_anomaly2} and \eqref{eq:trace_anom}.
This type of structure was studied in~\cite{Hinterbichler2023GravitySymmetries} and is sufficient to fix the $p^{-2}$ term in the $\ev*{X Y}$ correlator in any theory where this anomaly structure is present.
The set-up used in that work was slightly different as only a subset of the graviton shift symmetries of the form \eqref{eq:alpha=divLambda} were gauged and a [2,2] background gauge field was employed. However, it is clear that the methods developed there (see also \cite{Delacretaz:2019brr}) will apply equally here.
The result is that the non-local part of the correlator is precisely of the form \eqref{eq:XY_correlator}. This does not fix terms polynomial in $p^\mu$ in the correlator. In position space, these are contact terms and they depend on whether the anomaly appears in the non-conservation of $X$ or $Y$ when both currents are inserted at the same spacetime point.
The salient feature of the result is that the correlator has a $p^{-2}$ pole which dictates the IR behaviour of the correlator, and so indicates a massless mode in the IR spectrum. The results of \cite{Weinberg:2020nsn, Distler:2020fzr} then imply that this is necessarily a spin-two mode. 

Furthermore, in a theory with such an anomaly structure, the presence of the massless spin-two mode determines that the low-energy effective theory is the graviton theory with action \eqref{eq:S_FP} at leading order.\footnote{At higher order, only terms built from the linearised Riemann tensor and its derivatives will enter the EFT as these respect the global [1,1] bi-form symmetry.} In this IR gauge theory, one can then identify the [1,1] bi-form and $[d-3,1]$ bi-form shift symmetries and construct the associated (non-gauge-invariant) currents $J$ and $\tilde{J}$ from the graviton. 
From these, gauge-invariant currents and $\star_L R$ and $R$ can be constructed by \eqref{eq:X_def} and \eqref{eq:Y_def}. Then $X$ and $Y$ are proportional to $R$ and $\star_L R$ in the graviton theory that emerges in the deep IR.\footnote{At higher energies, the dual $[d-2,2]$ current $Y =\star_L R$ will remain unchanged as it is identically left-conserved, but the [2,2] current $X$ will be corrected by higher-order terms.}
Therefore, given the anomaly of $X$ and $Y$ is present in a given theory, the currents $J$ and $\tilde{J}$ can be constructed in the deep IR despite the fact that they are not gauge-invariant.
Furthermore, while $X$ and $Y$ were the observables defining the anomaly, their behaviour is completely defined by that of $J$ and $\tilde{J}$.
It is clear from this discussion that, while not gauge-invariant, the currents $J$ and $\tilde{J}$ contain all the physical information about the gauge-invariant currents $X$ and $Y$. They are, in this sense, the more fundamental quantities as they are the Noether currents associated with the most general global shift symmetries of the graviton and its dual. The gauge-invariant objects $X$ and $Y$ are objects derived from $J$ and $\tilde{J}$.

The form of the correlators resulting from the anomaly between $J$ and $\tilde{J}$ are consistent with those given in \cite{Hinterbichler2023GravitySymmetries}. However, understanding this anomaly from the more general graviton shifts provides a cleaner route to the anomaly and a more natural interpretation of the graviton as a Nambu-Goldstone field.

\section{Discussion \& outlook}
\label{sec:discussion}

We have found a mixed 't Hooft anomaly in the theory of the free graviton on a flat background spacetime. This is an obstruction to simultaneously gauging the shift symmetry of the graviton and the shift symmetry of the dual graviton. These are $[1,1]$ and $[d-3,1]$ bi-form symmetries respectively. The structure of the anomaly closely parallels the mixed 't Hooft anomaly present between the 1-form and $(d-3)$-form symmetries of Maxwell theory. 
The graviton can be viewed as a Nambu-Goldstone field for the shift symmetry \eqref{eq:graviton_shift} while the dual graviton can be viewed as a Nambu-Goldstone field for the dual shift symmetry \eqref{eq:D_shift}.

While the Noether currents $J$ and $\tilde{J}$ (defined in \eqref{JGam} and \eqref{eq:Jtilde}) associated with these shift symmetries are not gauge-invariant, we can 
construct gauge-invariant currents \eqref{eq:X_def} and \eqref{eq:Y_def} from these, which are proportional to the linearised Riemann tensor and its dual. The anomaly is reflected in the current algebra \eqref{eq:divX_contact} and we have related this anomaly to the one found in \cite{Hinterbichler2023GravitySymmetries} which implies that any theory with such an anomaly will have a massless graviton degree of freedom in the IR. 

Our derivation of the mixed 't Hooft anomaly avoids several complications present in \cite{Hinterbichler2023GravitySymmetries} which restrict their analysis to a subset of the most general shift symmetries of the graviton. In particular, the gauging presented here gives direct access to the current algebra of the symmetries and gives a more natural interpretation of the graviton as a Nambu-Goldstone boson. Furthermore, our presentation allows for the anomaly to be understood via the descent procedure with a simple anomaly polynomial \eqref{eq:anomaly_poly}, which closely parallels the higher-form mixed 't Hooft anomaly of Maxwell theory \eqref{eq:anomaly_poly_maxwell}.

To understand the quantum theory further, it is necessary to consider the observables of the theory.
We now briefly discuss observables in gravity. In General Relativity (GR), there are no local observables but there are extended observables. For an $n$-dimensional submanifold $\mathcal{N}$ of spacetime, the volume of $\mathcal{N}$ with respect to the spacetime metric $g$ is an observable, while for a curve $\mathcal{C}$ in spacetime one defines the gravitational holonomy, for example as the Wilson line operator constructed using the spin connection. For an asymptotically flat spacetime, one can also define the ADM charges \cite{ADM, Abbott1982StabilityConstant} as integrals at spatial infinity and the BMS charges \cite{Bondi:1962px, Sachs:1962wk} as integrals at null infinity.

For linearised gravity, in contrast, there are local observables: the linearised curvature and its derivatives are gauge invariant and so can be regarded as local observables, as can quantities constructed from functions of the curvature and its derivatives.

In this paper, we have focused on gravity linearised about Minkowski space, but much of our discussion generalises to gravity linearised about a spacetime with metric $\bar{g}_{\mu \nu}$; we will suppose that the background metric satisfies the (non-linear) Einstein equation. Linearising GR about this background gives a graviton field $h_{\mu \nu}$ satisfying Einstein's equations linearised about the background with gauge transformations in which \eqref{eq:linear_diffeo} is replaced by 
\begin{equation}\label{eq:nonlin_diffeo}
    \delta h_{\mu \nu} = \bar{\nabla}_{\mu} \xi_{\nu} + \bar{\nabla}_{\nu} \xi_{\mu}
\end{equation}
where $\bar{\nabla}$ is the Levi-Civita connection for $\bar{g}_{\mu \nu}$.
For an $n$-dimensional submanifold $\mathcal{N}$ of spacetime with (world-volume) coordinates $\sigma^a$, consider the quantity
\begin{equation}
    V_{\mathcal{N}} = \int_{\mathcal{N}} d^n \sigma \sqrt{- \bar{\gamma}} \,\bar{\gamma}^{a b} h_{\mu \nu} \pdv{x^{\mu}}{\sigma^a} \pdv{x^{\nu}}{\sigma^b}
\end{equation}
where $\bar{\gamma}_{a b}$ is the pull-back metric
\begin{equation}
    \bar{\gamma}_{a b} = \bar{g}_{\mu \nu} \pdv{x^{\mu}}{\sigma^a} \pdv{x^{\nu}}{\sigma^b}
\end{equation}
Now $V_{\mathcal{N}}$ is invariant under the gauge transformation \eqref{eq:nonlin_diffeo} provided that $\mathcal{N}$ is an extremal surface for the metric $\bar{g}_{\mu \nu}$, i.e. it is one that extremises the $n$-volume measured with respect to $\bar{g}_{\mu \nu}$; for a curve with $n = 1$ it is a geodesic. Then $V_{\mathcal{N}}$ can be
regarded as an observable associated with the extremal surface.

Whereas in the non-linear theory the ADM and BMS momentum and angular momentum charges are defined by integrals at infinity, in the linear theory there are momentum and angular momentum charges defined as   integrals over submanifolds of spacetime (which do not need to be at infinity) which define topological operators as they are unchanged under deformations of the submanifolds \cite{Benedetti2023GeneralizedGravitons, Hinterbichler2023GravitySymmetries, Hull:2024xgo}. Moreover there are further magnetic topological charges \cite{HullYetAppear} and gauge-invariant forms for these charges were found and discussed in \cite{Hull:2024mfb,Hull:2024xgo}.

It will be interesting to understand the full set of observables for linearised gravity and the action of the generalised symmetries of the theory on them. 
Here, we have focused on the symmetries of the graviton theory. In the context of the Landau paradigm, the observables charged under the symmetries will be order parameters which describe how the global symmetries are realised in a given phase (i.e. whether they are spontaneously broken or not). In four-dimensional Maxwell theory, the Wilson line is the order parameter for the electric 1-form symmetry. In a Coulomb phase it has perimeter law behaviour and the symmetry is spontaneously broken, with the photon arising as a Nambu-Goldstone boson. It would be interesting to identify the order parameters in gravity whose long-distance behaviour would describe the spontaneously breaking of the bi-form symmetries in the linear theory.

There has been extensive work on $p$-form generalised symmetries in the past decade. Our results (together with those of \cite{Hinterbichler2023GravitySymmetries, Hinterbichler:2024cxn}) indicate that much of the formalism and structure associated with higher-form symmetries generalises to symmetries with certain bi-form parameters.
Further generalisations to general symmetries in which the gauge parameters are in arbitrary Lorentz representations should be straightforward and would have applications to higher spin gauge fields in arbitrary Lorentz representations. 
Such symmetries can be regarded as \emph{multi-form} symmetries.\footnote{That is, the parameter of a general shift symmetry will be a multi-form, which is a tensor with multiple sets of anti-symmetric indices. The multi-form calculus of \cite{Medeiros2003ExoticDuality} will be useful here.}
However, one important difference with the higher-form symmetries is the following.
A $p$-form current can be integrated over a $p$-surface to give a topological operator, but a multi-form current cannot be directly integrated in this way.
However, as we have discussed, topological operators can be defined by contracting a multi-form current with a suitable tensor to yield a $p$-form and then integrating. A given multi-form symmetry can in general produce a number of different higher-form currents by contracting with different tensors.

The symmetries of the linear graviton theory discussed here do not appear to have a generalisation to the non-linear Einstein theory. Of course, a quantum theory of gravity is expected to have no global symmetries \cite{Banks:2010zn, Harlow:2018tng}, but this is not the case for the linearised quantum theory. The study of this system led us to currents corresponding to these symmetries and to an understanding of their 2-point correlation functions.
This was then used to study the possibility of the emergence of a graviton in a matter system, following arguments in \cite{Hinterbichler2023GravitySymmetries}.
We argued that an arbitrary matter system with a pair of currents with the properties and 2-point function discussed here must necessarily have a gapless (massless) spin-two state. 
It would be interesting to understand whether an interacting graviton could emerge in such a scenario.
The possibility of the emergence of gravity from a non-gravitational system is of considerable interest; see e.g.~\cite{Verlinde:2016toy}.

\paragraph{Acknowledgements.}
CH is supported by the STFC Consolidated Grants ST/T000791/1 and ST/X000575/1.
MLH is supported by a President's Scholarship from Imperial College London.
UL gratefully acknowledges a Leverhulme Visiting Professorship to Imperial College as well as the hospitality of the theory group at Imperial.

\appendix
\section*{Appendix}

\section{Quantisation \& charges generating shift symmetries}
\label{app:quantisation}

In this appendix we canonically quantise the graviton theory and demonstrate that the charge $Q(\alpha)$ in \eqref{eq:Q(alpha)} does, indeed, generate a shift of the graviton field. 

The graviton action is given in \eqref{eq:S_FP}. We decompose the Minkowski space coordinates as $x^\mu = (x^0\equiv t, x^i)$ with $i=1,\dots,d-1$. Integrating the action by parts allows time derivatives to be shuffled between different terms. Doing so then changes the canonical momenta $\pi^{00}$, $\pi^{0i}$, and $\pi^{ij}$. We will take the convention used, for example, in \cite{Hinterbichler:2011tt} where all time derivatives in the action are removed from the fields $h_{00}$ and $h_{0i}$ by integration by parts.
These fields then appear as Lagrange multipliers, enforcing $d$ first class constraints.
The dynamical fields are then the $h_{ij}$, whose canonical momenta are
\begin{equation}
    \pi^{ij} = \frac{1}{2} \left( \dot{h}_{ij} - \dot{h} \delta_{ij} + 2\partial_k h_{k0} \delta_{ij} - 2\partial_{(i} h_{j)0} \right)
\end{equation}
where $\dot{f} \equiv \partial_0 f$ and $h \equiv h\indices{_i^i}$. 

Recall that the charge $Q(\alpha)$ in \eqref{eq:Q(alpha)} is the charge associated with the conserved 1-form current $j(\alpha)_\mu$ in \eqref{eq:j(alpha)}. 
Now, taking the surface $\Sigma_{d-1}$ on which it is defined to be a constant-time hypersurface at time $t$, $\Sigma_{d-1} = \mathbb{R}^{d-1}_t$, the charge is
\begin{equation}
    Q(\alpha) = \int_{\Sigma_{d-1}} \dd[d-1]{\vec{x}} j_0(\alpha)
\end{equation}
where $\vec{x} = (x^1,\dots,x^{d-1})$.
Conservation implies that $Q(\alpha)$ is independent of $t$.

Expanding the current explicitly in the $(t,x^i)$ coordinate split, the charge can be written
\begin{equation}
    Q(\alpha) = \int \dd[d-1]{\vec{x}} \left( \alpha_{ij} \pi_{ij} - \alpha_{i0} \left( \partial_j h_{ij} - \partial_i h \right) - \frac{1}{2} \dot{\alpha}_{ij} \left( h_{ij} - \delta_{ij} h \right) \right)
\end{equation}
where we assume that the field vanishes sufficiently fast at spatial infinity that we can drop boundary terms.\footnote{We could also consider closed $\Sigma_{d-1}$.}

We may now pass directly to the quantum theory by replacing Poisson brackets by $i\hbar$ times canonical equal-time commutators to obtain (with $\hbar=1$)
\begin{equation}
    \comm{h_{ij}(\vec{x})}{\pi^{kl}(\vec{y})} = i \delta_{(i}^k\delta_{j)}^l \delta(\vec{x}-\vec{y}) 
\end{equation}
which imply that the action of $Q(\alpha)$ on the dynamical fields $h_{ij}$ is
\begin{align}
    \comm{h_{ij}(\vec{x})}{Q(\alpha)} &= i \alpha_{ij}(\vec{x})
\end{align}
provided that $\vec{x} \in \Sigma_{d-1}$, and otherwise the result vanishes vanishes. Therefore, the charge $Q(\alpha)$ generates the correct shifts of the graviton field in this canonical formulation.\footnote{We emphasise that the $h_{00}$ and $h_{0i}$ fields are unphysical in this formulation and so $Q(\alpha)$ is the correct operator to generate shifts of the \emph{physical} fields.}

The algebra of these charges can be computed using the commutator above. A straightforward but lengthy calculation gives 
\begin{equation}
\begin{split}\label{eq:comm_working}
    \comm{Q(\alpha)}{Q(\alpha')} &= \frac{i}{2} \int \dd[d-1]{\vec{x}} \big( 2 \alpha_{i0} \partial_i \alpha' - 2\alpha_{i0} \partial_j \alpha'_{ji} - 2\partial_i \alpha \alpha'_{i0} + 2\partial_j \alpha_{ji} \alpha'_{i0}\\
    & \qquad\qquad\qquad + \alpha_{ij} \dot{\alpha}'_{ij} - \alpha \dot{\alpha}' - \alpha'_{ij} \dot{\alpha}_{ij} + \alpha' \dot{\alpha} \big)
\end{split}
\end{equation}
where $\alpha \equiv \alpha_{ii}$, $\alpha' \equiv \alpha'_{ii}$ and we have assumed that $\alpha$, $\alpha'$ vanish sufficiently fast at the boundary of $\Sigma_{d-1}$, so we can integrate the spatial derivatives by parts.
This result can be neatly written in terms of $\dd_L \alpha$ and $\dd_L \alpha'$, which have components $(\dd_L \alpha)_{\mu\nu|\rho} = \partial_{[\mu} \alpha_{\nu]\rho}$. This implies the relations
\begin{equation}
\begin{split}
    2(\dd_L \alpha)_{ij|j} &= \partial_i \alpha - \partial_j \alpha_{ij} \\
    2(\dd_L \alpha)_{0i|j} &= \dot{\alpha}_{ij} - \partial_i \alpha_{j0} \\
    2(\dd_L \alpha)_{0ii} &= \dot{\alpha} - \partial_i \alpha_{i0}
\end{split}
\end{equation}
Substituting these into \eqref{eq:comm_working} and integrating spatial derivatives by parts gives
\begin{equation}
    \comm{Q(\alpha)}{Q(\alpha')} = i \int\dd[d-1]{\vec{x}} \left( 3 \alpha\indices{^i_{[0}} (\dd_L\alpha')\indices{_{ij]|}^j} - (\alpha\leftrightarrow\alpha') \right)
\end{equation}
The integrand can be recognised as the $t$-component of the 1-form $\chi(\alpha,\alpha')$ defined in \eqref{eq:zeta_def},
such that the whole equation can be written in a Lorentz covariant form
\begin{equation}
    \comm{Q(\alpha)}{Q(\alpha')} = i \int_{\Sigma_{d-1}} \star\chi(\alpha,\alpha')
\end{equation}

\bibliographystyle{JHEP}
\bibliography{references}

\end{document}